\documentclass[]{aa}
\usepackage{natbib}
\usepackage{graphicx}
\usepackage{color}
\usepackage{amsmath}
\usepackage{url}
\usepackage{amssymb}

\newcommand\bb[1] {   \mbox{\boldmath{$#1$}}  }
\newcommand\del{\bb{\nabla}}
\newcommand\bcdot{\bb{\cdot}}
\newcommand\btimes{\bb{\times}}

\begin{document}

\title{Vortex cycles at the inner edges of dead zones
in protoplanetary disks}
\author{Julien Faure \inst{1}
\and S\'ebastien Fromang \inst{1} 
\and Henrik Latter \inst{2}  
\and Heloise Meheut \inst{1}}

\offprints{J.Faure}

\institute{Laboratoire AIM, CEA/DSM--CNRS--Universit\'e Paris 7, Irfu/Service d'Astrophysique, CEA-Saclay, 91191 Gif-sur-Yvette, France \and Department of Applied Mathematics
and Theoretical Physics, University of Cambridge, Centre for
Mathematical Sciences, Wilberforce Road, Cambridge, CB3 0WA, UK\\
\email{julien.faure@cea.fr}}

\date{Accepted/ Received/ in original form}

\abstract{In protoplanetary disks, the inner boundary between the
  turbulent and laminar regions is a promising site for planet
  formation because solids may become trapped
  at the interface itself or
  in vortices generated by the Rossby wave instability. The disk
  thermodynamics and the turbulent dynamics at that location are
  entwined because of the importance of turbulent dissipation on
  thermal ionization and, conversely, of thermal ionisation on the
  turbulence. 
  However, most previous work has
  neglected this dynamical coupling and have thus missed a key element of
  the physics in this region.}
{In this paper, we aim to determine how the the interplay
  between ionization and turbulence impacts on the formation and
  evolution of vortices at the interface between the active and the
  dead zones.} 
{Using the Godunov code RAMSES, we have performed a 3D magnetohydrodynamic global
  numerical simulation of a cylindrical model of an MRI--turbulent
  protoplanetary disk, including 
  thermodynamical effects as well as a temperature-dependant
  resistivity. The comparison with an analogous 2D viscous simulation
  has been extensively used to help identify the relevant physical
  processes and the disk's long-term evolution.}
{We find that a vortex formed at the interface, due to
 Rossby wave instability, migrates inward and penetrates the active zone
 where it is
  destroyed by turbulent motions. Subsequently, a new vortex emerges a
  few tens of orbits later at the interface, and the new vortex
  migrates inward too. The sequence repeats itself, resulting
   in cycles of vortex formation, migration, and disruption. This
   surprising behavior is successfully reproduced using two different
   codes.
  In this paper, we characterize
  this vortex life cycle and discuss its 
  implications for planet formation at the dead/active interface.
  Our results also call for a better understanding of vortex
  migration in complex thermodynamical environments.}
{Our simulations highlight the importance of thermodynamical processes
  for the vortex evolution at the dead zone inner
  edge.}


\maketitle

\authorrunning{J.Faure et al.}
\titlerunning{Vortex dynamics}

\section{Introduction}

In current planet formation theory, the agglomeration of microscopic dust
into kilometre-size objects remains poorly understood
\citep{2010AREPS..38..493C}.  
The direct gravitational collapse scenario \citep{1973ApJ...183.1051G, 1980Icar...44..172W, 1981Icar...45..517N, 1985prpl.conf.1100H}
requires a midplane dust-to-gas ratio larger than the
interstellar medium value by about three orders of magnitude,
which is not easily achievable by sedimentation alone 
\citep{1993prpl.conf.1031W, 1998Icar..133..298S, 2010ApJ...708..188T}.
Growth resulting from ballistic collisions and sticking works
  efficiently for small particles
\citep{1993ApJ...407..806C, 1997ApJ...480..647D, 2000ApJ...533..472P},
but laboratory and numerical experiments suggest two
limiting grain sizes ($\sim 1 mm$ at 1 AU), known as the bouncing and fragmentation
barriers, beyond which colliding particles do not stick
\citep{1993prpl.conf.1031W, 2000SSRv...92..279B,
  2010ApJ...725.1242K, 2005Icar..178..253W,
  2010A&A...513A..57Z, 2012A&A...544L..16W,
  2013A&A...551A..65S, 2013ApJ...764..146G}.
Note however the recent calculations of the evolution of a
  dust population presented by \citet{2013ApJ...764..146G}, who argue
  these barriers are less insurmountable than first thought.
  For example, incorporation of a realistic particle velocity
  distribution ameliorates the problem, as does
  the realisation that high--mass-ratios collisions can transfer
  mass to the largest of two impactors.

A separate problem is that the solid content of a protoplanetary (PP) disk rapidly
drains out of the disk
\citep{1997A&A...319.1007S,2001A&A...378..180K, 
  2002ApJ...581.1344T,2007A&A...469.1169B,2012MNRAS.423..389H}. This is because the gas is 
partly supported by the disk's pressure gradient and rotates at slightly
sub-Keplerian frequencies, while dust grains
rotate faster at the Keplerian angular velocity.  
Consequently, dust grains feel a `head-wind' and thereby lose their angular momentum,
ultimately spiralling into the central star. The
inward drift velocity is maximal for particles of a few tens of
centimetres and such particles only survive for some 100 years \citep{1977MNRAS.180...57W}. 
This phenomenon constitutes an additional barrier known as the
radial-drift barrier. 

Among the many scenarios that have been discussed to
overcome these problems, some of the most promising involve the
pressure bumps formed at planetary gap edges
\citep{2007A&A...471.1043D,2011MNRAS.415.1426L} or 
at the interface between a laminar (`dead') and a turbulent (`active') region
\citep{2007ApJ...664L..55K,2008A&A...487L...1B,kretkeetal09}. Friction
between gas and dust particles vanishes at the location of the
pressure maximum, naturally providing a trap where the disk's solid
content can accumulate. In addition, such pressure extrema are unstable
to the vortex or Rossby wave instability 
\citep[RWI,][]{1999ApJ...513..805L,2000ApJ...533.1023L,2010A&A...516A..31M,2012ApJ...754...21L}.
It is well established that vortices significantly concentrate
dust in a short time 
\citep{barge&sommeria95,1996Icar..121..158T,2004A&A...417..361J,2006ApJ...639..432K,2006ApJ...649..415I,meheutetal12,2012A&A...542A...9M},
thus promoting planet formation at these locations. We finally note
that the presence of vortices has been suggested by high
angular resolution imaging of PP disks which exhibit
significant asymmetric features
\citep{2009ApJ...704..496B,2012ApJ...754L..31C,2013Sci...340.1199V,2013ApJ...775...30I,2014ApJ...783L..13P}. 

In this context, the dead-zone inner edge of PP disks is a 
special location that deserves detailed investigation. We define
the inner edge to be the radius where the (radially decreasing) midplane
temperature drops below the critical
value at which thermal ionization fails to sustain the
magnetorotational instability \citep[MRI, ][]{1991ApJ...376..214B,1998RvMP...70....1B}. 
As a result, the flow is turbulent inward of that interface but laminar
beyond it \citep{1996ApJ...457..355G}. Recent 3D MHD simulations
have reported the 
formation of pressure bumps in
locally isothermal models of PP disks \citep{nataliaetal10,lyra&maclow12},
confirming earlier 2D hydrodynamical viscous simulations.
\citep{varniere&tagger06,2008A&A...491L..41L,2012MNRAS.419.1701R}. 
This work has hence established the feasibility of the 
dead-zone inner edge as a trap for solids, however the robustness of
these results is
curtailled by the use of simple thermal physics, 
namely a locally isothermal equation of
state. 
This assumption is problematic because of the pervasive
interpenetration of dynamics and thermodynamics in this
region, especially at the midplane. Temperature depends on the
turbulence via the dissipation of its kinetic and magnetic fluctuations,
but the MRI turbulence, in turn, depends on the temperature
through the ionization fraction, which is determined
thermally via dissipation (Pneuman \& Mitchell 1965; Umebayashi \&
Nakano 1988). In addition, the onset and nonlinear evolution of the
RWI should depend on the PP disk's global thermal structure through its
radial potential-vorticity profile (Umurhan 2010).

Using a simplified mean
field model, \citet{latter&balbus12} found that, if the interplay
between thermal and turbulent dynamics is taken into account,
 the dead/active interface is not static but rather moves radially
 before stalling at a well-defined radius.
 Recently, we confirmed this behaviour using
MHD simulations that
self-consistently accounted for both turbulent heating and the feedback
between temperature and magnetic diffusivity \citep{MOI}. However, in
order to reduce the computational cost of these simulations, vortex
formation was artificially inhibited by using a reduced azimuthal
domain. 
The point of the present paper is
to examine the onset and development of vortices in thermally
structured models of PP disks. We do so by increasing the azimuthal extent of our
previous simulations.

In order to isolate and understand the basic physics we adopt
simplified geometry and thermodynamics. Our PP disk is
cylindrical, and hence vertically unstratified. Consequently, disk
cooling is approximated by a cooling law rather than via a detailed
radiative transport model. As expected, we find that a vortex forms at the dead zone
inner edge via the RWI, but contrary to expectations (e.g.\
Paardekooper et al.~2010) the vortex
radially migrates inwards, ploughing through the pressure bump and
into the active zone where it is ultimately destroyed by turbulent
motions. A few orbits later a new vortex forms at the pressure bump
and it too follows the same cycle of formation, migration, and disruption.
This behaviour fails to appear in isothermal or adiabatic runs, and
seems connected to the details of the PP disk's heating and cooling. 
It is yet unclear how robust this `vortex cycle' is, in particular how
sensitive it is to the approximate cooling law we have employed. Future
vertically structured global simulations using the set-up of
Flock et al.~(2013) will aid in testing this. Taken on face value,
however, these results complicate planet formation theories that
appeal to dust trapping at the inner dead-zone edge unless a
fast formation process is at work within vortices.

The paper is organized as follows. In Sect.~\ref{3D} we present
the results of a 3D MHD simulation which exhibits
 the basic vortex cycle. 
Sect.~\ref{2D} contains the results of a 2D run of a simple viscous model
that reproduces the vortex behaviour observed in the 3D simulation.
This simplified model is a useful tool with which 
to analyse the vortex cycle in more detail. In
Sect.~\ref{disc}, we discuss the physical mechanisms
at work in both simulations. Finally, we conclude and speculate on the
implications of our simulations for planet formation in
Sect.~\ref{ccl}. 

\section{The 3D turbulent MHD simulation}
\label{3D}

We first present results from a 3D simulation subject to MRI-induced
turbulence. We focus on an annular region centred around
the dead zone inner edge, as in \citet{MOI}, to which the reader is
referred for further details. However, by extending the azimuthal
extend of the domain we observe the onset and development of the RWI. 
The behaviour of the vortices so formed is described in this section and
comprise the main result of the paper. 

\subsection{Setup}
The 3D simulation has been performed using a uniform grid version of
the code RAMSES \citep{teyssier02, fromangetal06}. We solve the MHD
equations in cylindrical coordinates $(R,\phi,Z)$ with units vectors
$(\bb{e_R},\bb{e_\phi},\bb{e_z})$:
\begin{eqnarray}
&\frac{\partial \rho}{\partial t}+\del \bcdot (\rho \bb{v}) = 0 \\
&\frac{\partial \rho \bb{v}}{\partial t}+\del \bcdot (\rho \bb{v} \bb{v} - \bb{B} \bb{B}) + \del P =
-\rho \del \Phi \\
&\frac{\partial E}{\partial t}+\del \bcdot \left[ (E+P) \bb{v}
- \bb{B} (\bb{B} \bcdot \bb{v}) + {\cal F_{\eta}} \right] = -\rho\bb{v}\cdot\nabla \Phi
-{\cal L} \label{energy_eq2} \\
&\frac{\partial \vec{B}}{\partial t} - \bb{\nabla} \btimes
(\bb{v} \times \bb{B}) = - \bb{\nabla} \btimes (\eta \bb{\nabla}
\btimes \bb{B}) \label{induc}
\end{eqnarray}
where $\rho$ is the density, $\bb{v}$ is the velocity, $\bb{B}$ is the
magnetic field, $P$ is the pressure and $E$ is the sum of kinetic, magnetic, and thermal energy. 
$\Phi$ is the gravitational potential.
In the cylindrical approximation, it is given by
$\Phi=-GM_{\star}/R$ where $G$ is the gravitational constant and
$M_{\star}$ is the stellar mass. 
We use a perfect gas equation of state to close the former set
of equations. The thermal energy is related to the pressure through the
relation $e_{th}=P/(\gamma-1)$ in which $\gamma=1.4$. The magnetic
diffusivity is denoted by $\eta$ and is responsible for the resistive
flux $\cal F_{\eta}$ that appears in Eq.~(\ref{energy_eq2})
\citep[see][]{1998RvMP...70....1B}. We capture turbulent 
heating by solving the equation of total energy conservation and use
the same gas cooling function $\mathcal{L}$ as in \citet{MOI}: 
\begin{equation}
{\cal L} =\rho \sigma (T^4-T_\text{min}^4) \, ,
\label{thermo_sec_eq}
\end{equation}
where $T$ is temperature, and $T_\text{min}$ is the temperature
associated with radiative equilibrium. 
We model the rapid variation of the resistivity $\eta$ with the temperature by a step function:
\begin{equation}
\label{resist}
\eta(T)=
\begin{cases}
  \eta_0 \, \, \, \, \textrm{if} \, \, \, T<T_\text{MRI}\\
  0 \, \, \, \, \, \textrm{otherwise} \, ,
\end{cases} 
\end{equation}
where $T_\text{MRI}$ is the activation temperature for the MRI,
typically $\sim 1000$ K.

The initial magnetic field configuration is purely toroidal. Its
vertical profile is such that the integrated magnetic flux through a
vertical slice of the disk vanishes.  The computational domain is $R
\in [R_0,8 \, R_0]$, $\phi \in [0,\pi/2]$ and $Z \in[-0.3 \, R_0,0.3 
\, R_0]$ and has a resolution of $[320,160,80]$. 

The main
difference with the setup of \citet{MOI} is the removal of the
source term in the continuity equation. We found that the
term could interfere with the development of density features and hence
with the RWI. In addition, we tuned the viscosity's
radial profile in the buffer zones in order to avoid accretion
discontinuities at the buffer edges.
We give here for completeness the functional form of the viscosity profile we used:
\begin{equation}
\nu_{\textrm{buff}}=\alpha_{\textrm{buff}} \left(\frac{R}{R_{\textrm{buff}}}\right)^{-0.5} c_s H
\end{equation}
where $\alpha_{\textrm{buff}}$ is the value of $\alpha$ measured at
$R=R_{\textrm{buff}}$ and averaged since the beginning of the
simulation. $R_{\textrm{buff}}$ is the position of the boundary
between the buffer zone and the active domain.
Together with free-flow boundary conditions this technique prevents unphysical
mass depletion in the domain.
 
\subsection{Notation and units}

In the following we denote by $X_0$ the value of any quantity $X$ at
the inner edge of the domain. Units are identical to that defined in
\citet{MOI} and chosen such that:
\begin{equation*}
GM_{\star}=R_0=\Omega_0=\rho_0=T_0=1 \, ,
\end{equation*}
where $\Omega$ stands for the gas angular velocity at radius $R$,
 and $\rho$ is density. 
Time is measured in units of the inner orbital period.

Density and temperature profiles are initialized with radial power laws:
\begin{equation} 
\rho=\rho_0 {\left(\frac{R}{R_0}\right)}^{p},  \quad
T=T_0 {\left(\frac{R}{R_0}\right)}^{q} \label{initT} \,,
\end{equation}
where $p=-1.5$ and $q=-0.75$. 

In the cooling law, equation \eqref{thermo_sec_eq}, we choose $\sigma=1.1 \times
10^{-4}$. For a typical simulation, this yields a disk aspect ratio of
 $H_0/R_0 \sim 0.1$ and a
 disk cooling time of about $25$ local orbits. Finally, we used 
$\eta_0=10^{-3}$, which is large enough to prevent the development of
the MRI in resistive regions, and
$T_{\textrm{MRI}}=0.6$.

\subsection{Results}

We force the simulation to undergo three consecutive phases, with a different
resistivity configuration in each. This approach introduces the
requisite physics in a controlled fashion. We describe each step in turn. 

\subsubsection{The ideal phase}

Initially the resistivity is set to zero
everywhere except in the buffer zones. The aim is to obtain a fully turbulent
disk free of unnecessary transient behaviour associated with the
initialisation of the MRI. Once this is achieved then we introduce a
dead zone in the following phase. 
The model
equations are integrated for a few cooling times and until the temperature reaches a
quasi-steady state. The radial profile of the equilibrium
temperature that results is plotted in figure~\ref{tempfull_turb}.

\subsubsection{The static dead-zone phase}

Once the simulation has reached a quasi-steady equilibrium ($t=600$) we set 
$\eta=\eta_0$ over the region $3.5<R<8$
and integrate the equations for another $500$ orbits when thermal
equilibrium is reached. MHD turbulence quickly dies outward of $R=3.5$ and
a static dead zone forms. Gas cools in this region as a result
of the absence of any turbulent heating. Soon we notice the formation of
density and pressure maxima around the interface at $R=3.5$, because
of the accretion mismatch, and this saturates
when the density reaches about twice its initial value. In a
few tens of orbits a vortex forms (see the first panel of
figure~\ref{vortex1}) at this location because of the RWI.

\subsubsection{The self-consistent dead-zone phase}

\begin{figure}
\begin{center}
\includegraphics[scale=0.4]{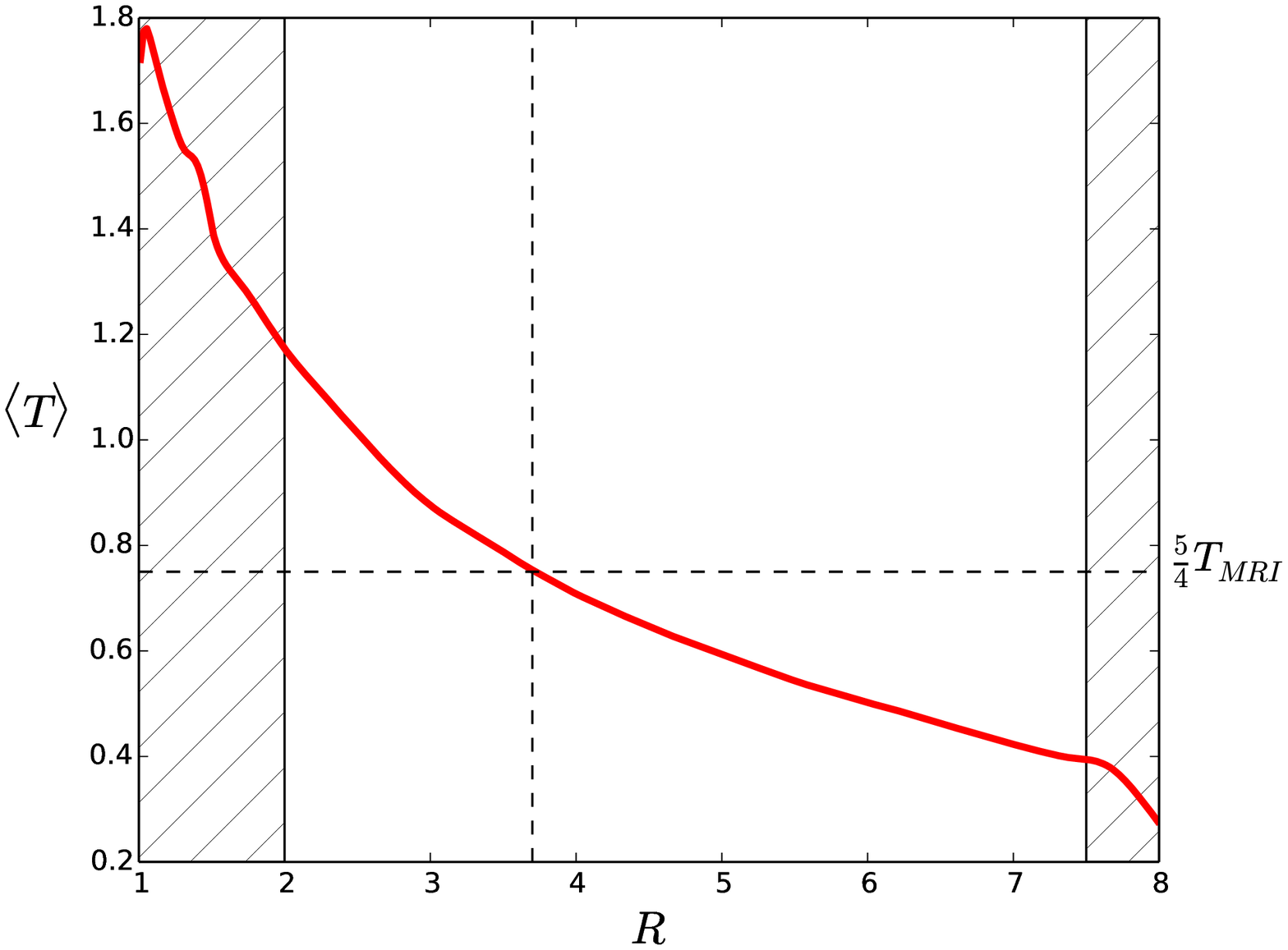}
\caption{Temperature profile at the end of the ideal phase
 averaged over $200$ inner orbits. The black dashed lines show where
 the mean temperature is equal to $(5/4) T_{\textrm{MRI}}$. The hashed
regions on both sides of the figure corresponds to the buffer zones.}  
\label{tempfull_turb}
\end{center}

\end{figure}
\begin{figure*}
\begin{center}
\includegraphics[scale=0.35]{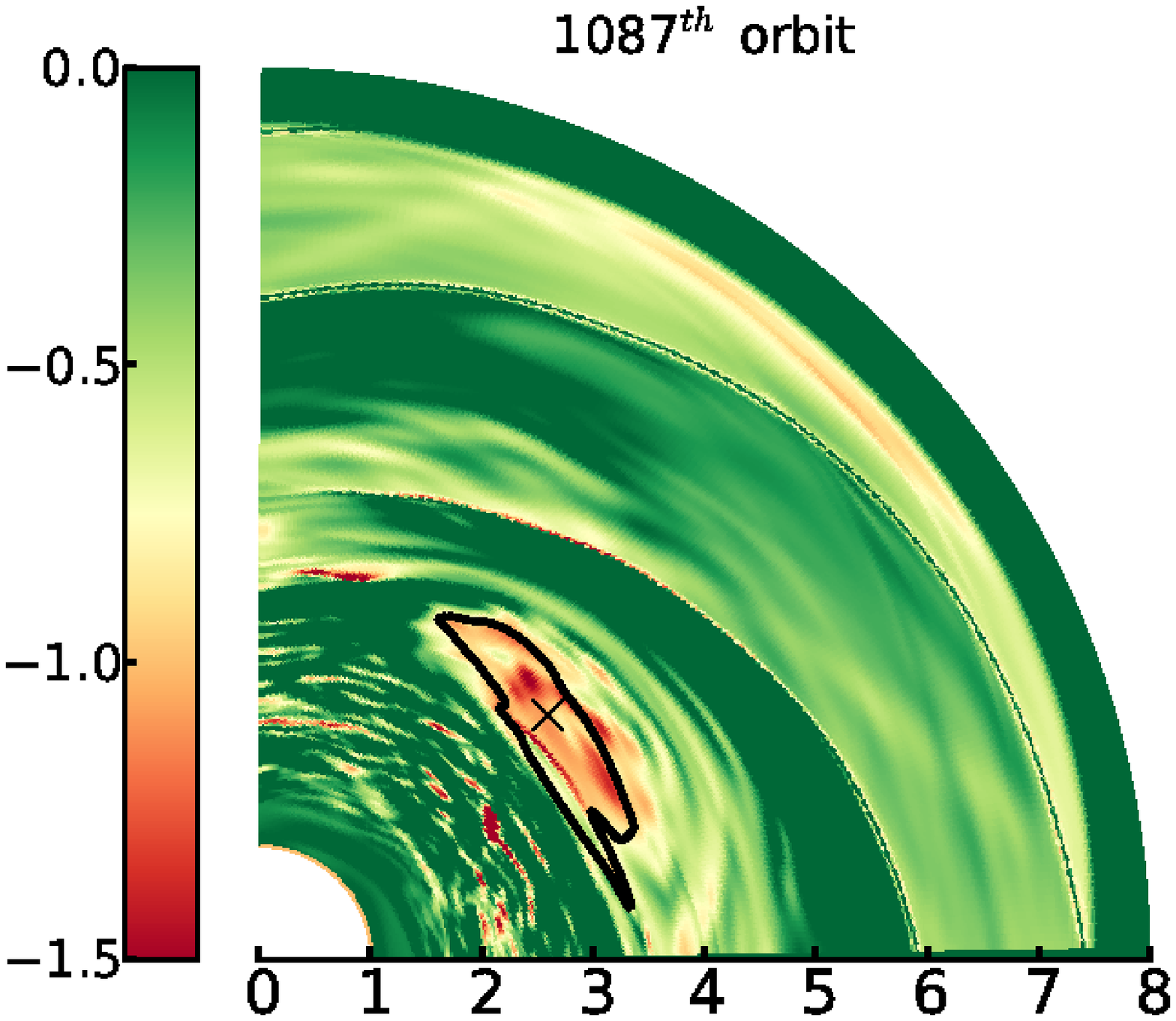}
\includegraphics[scale=0.35]{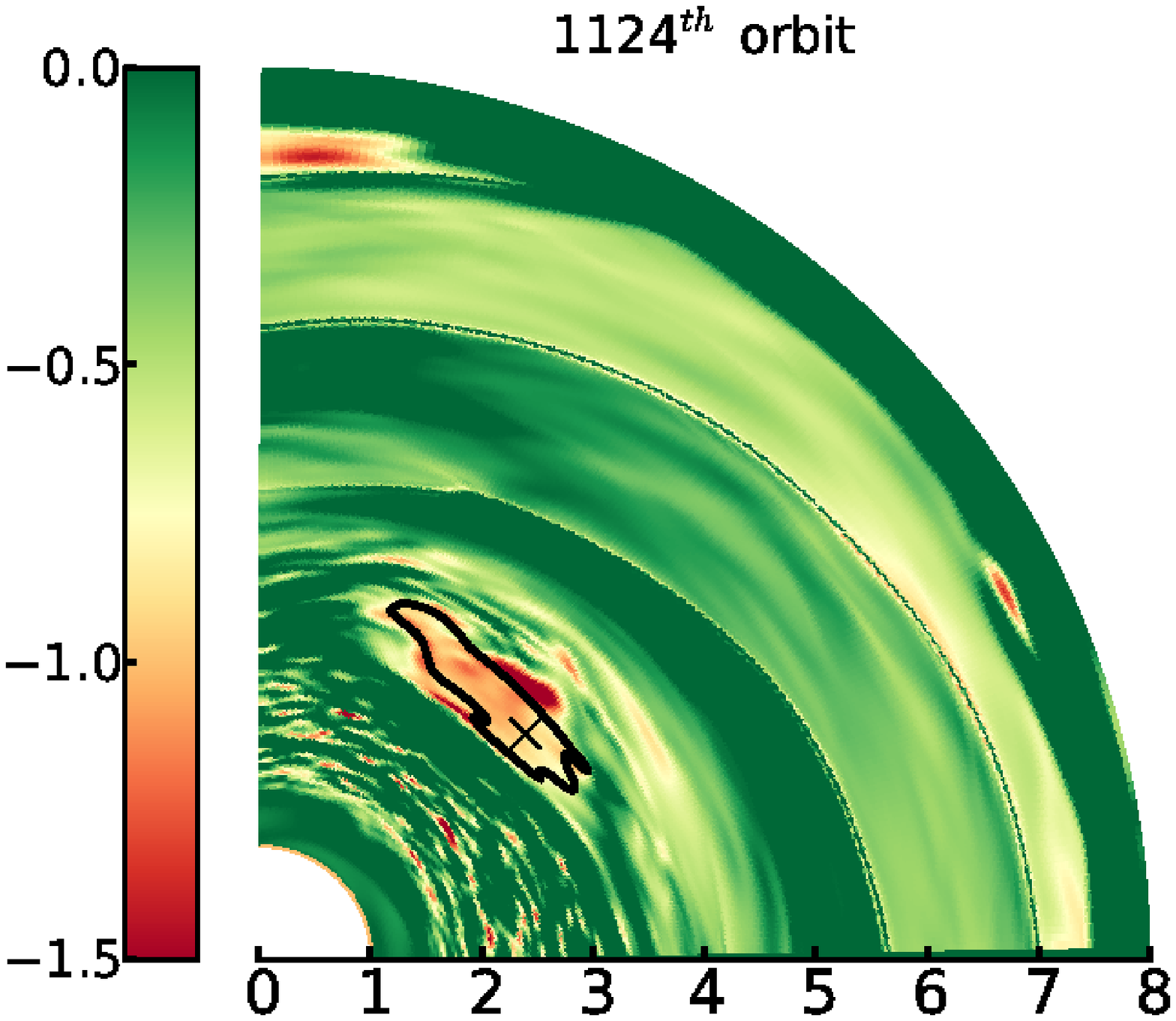}
\includegraphics[scale=0.35]{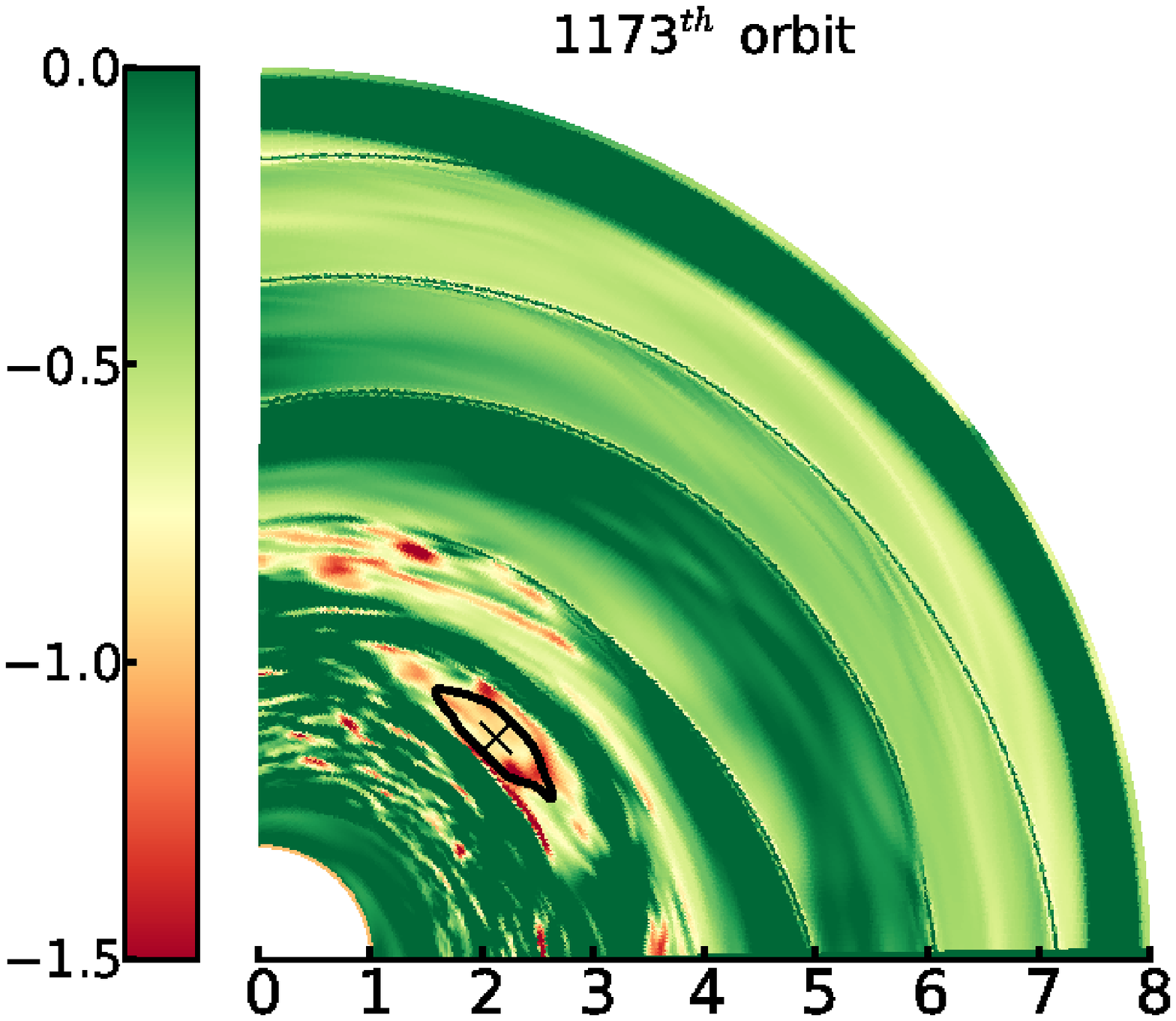}
\includegraphics[scale=0.35]{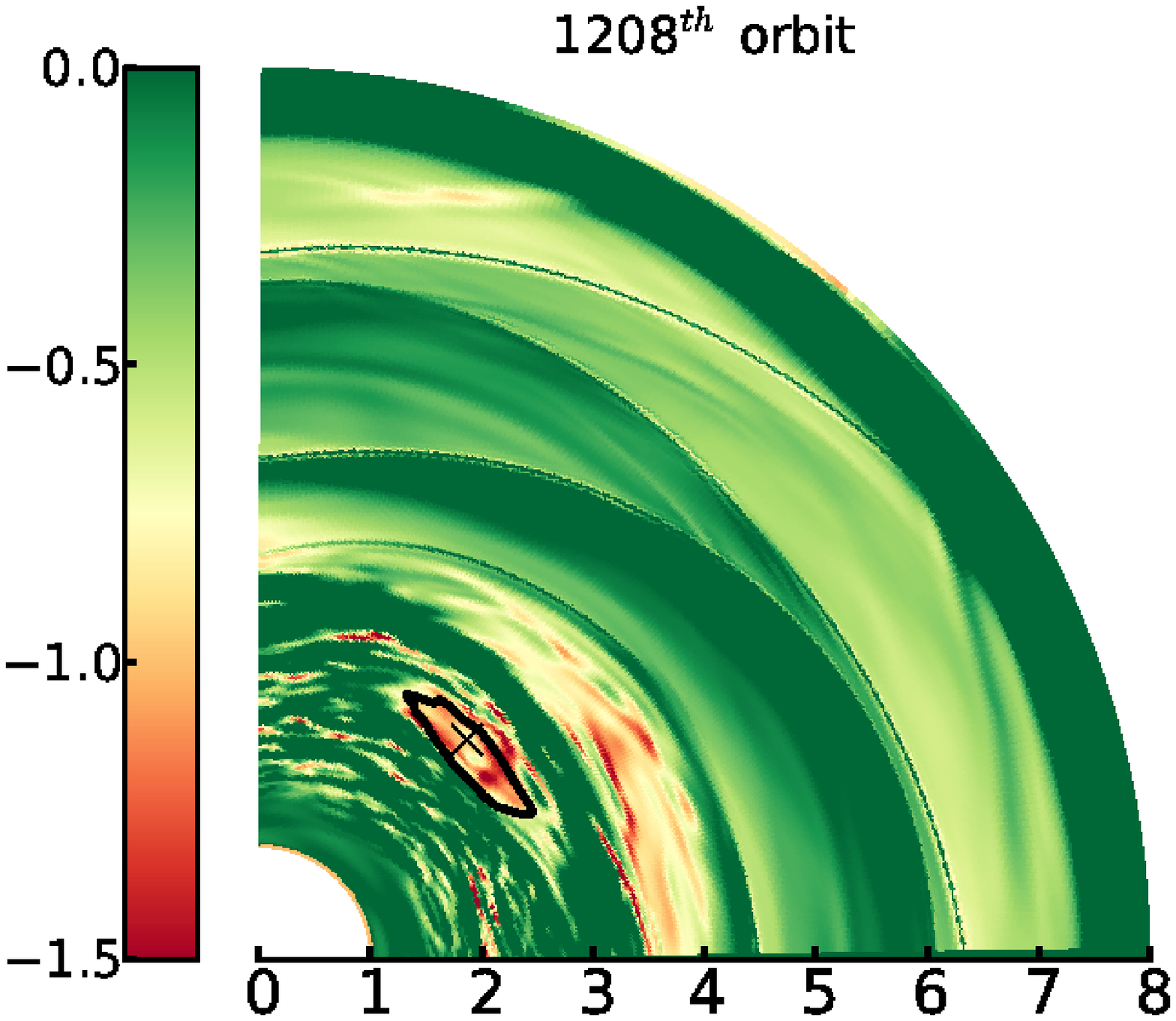}
\includegraphics[scale=0.35]{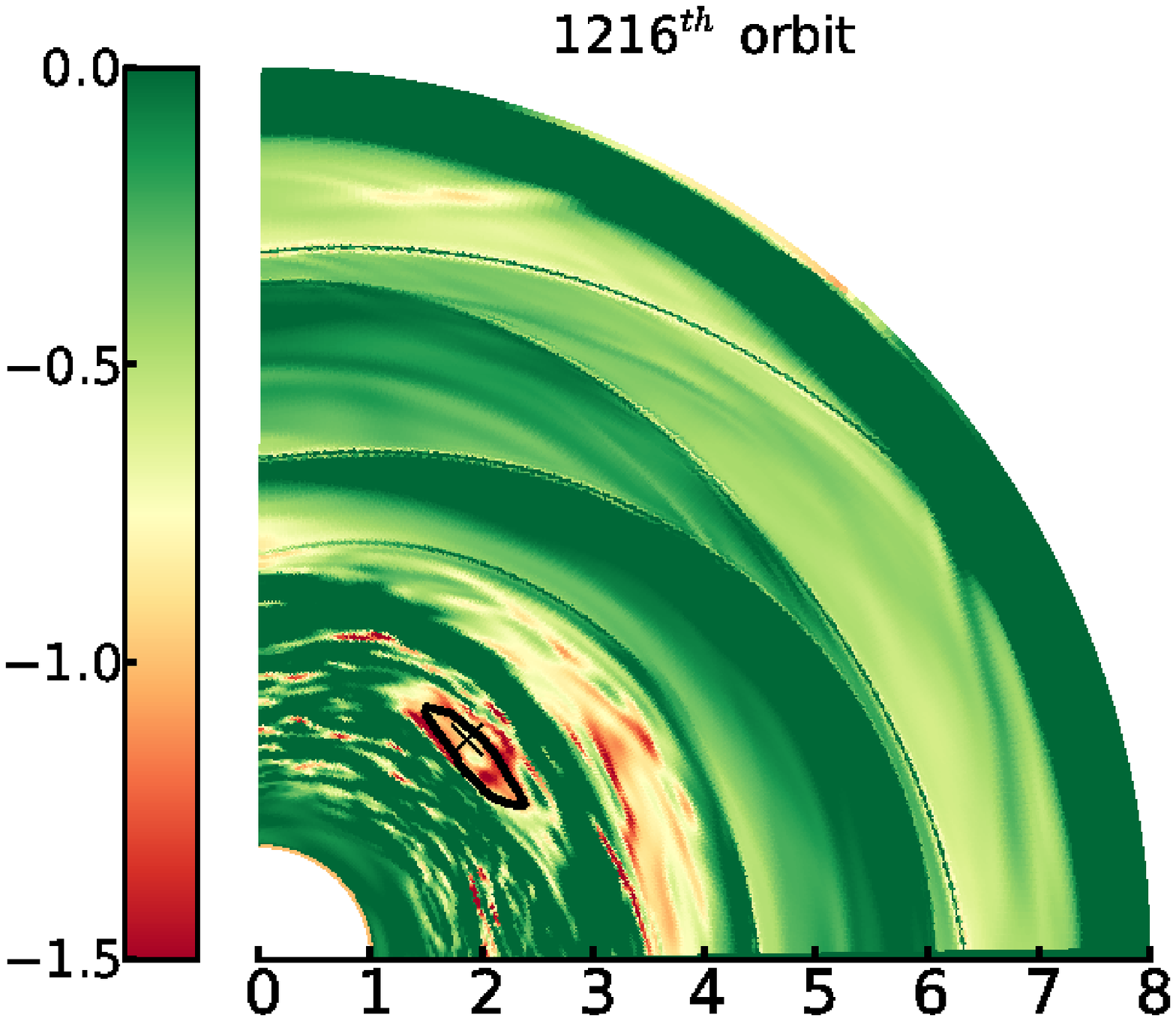}
\includegraphics[scale=0.35]{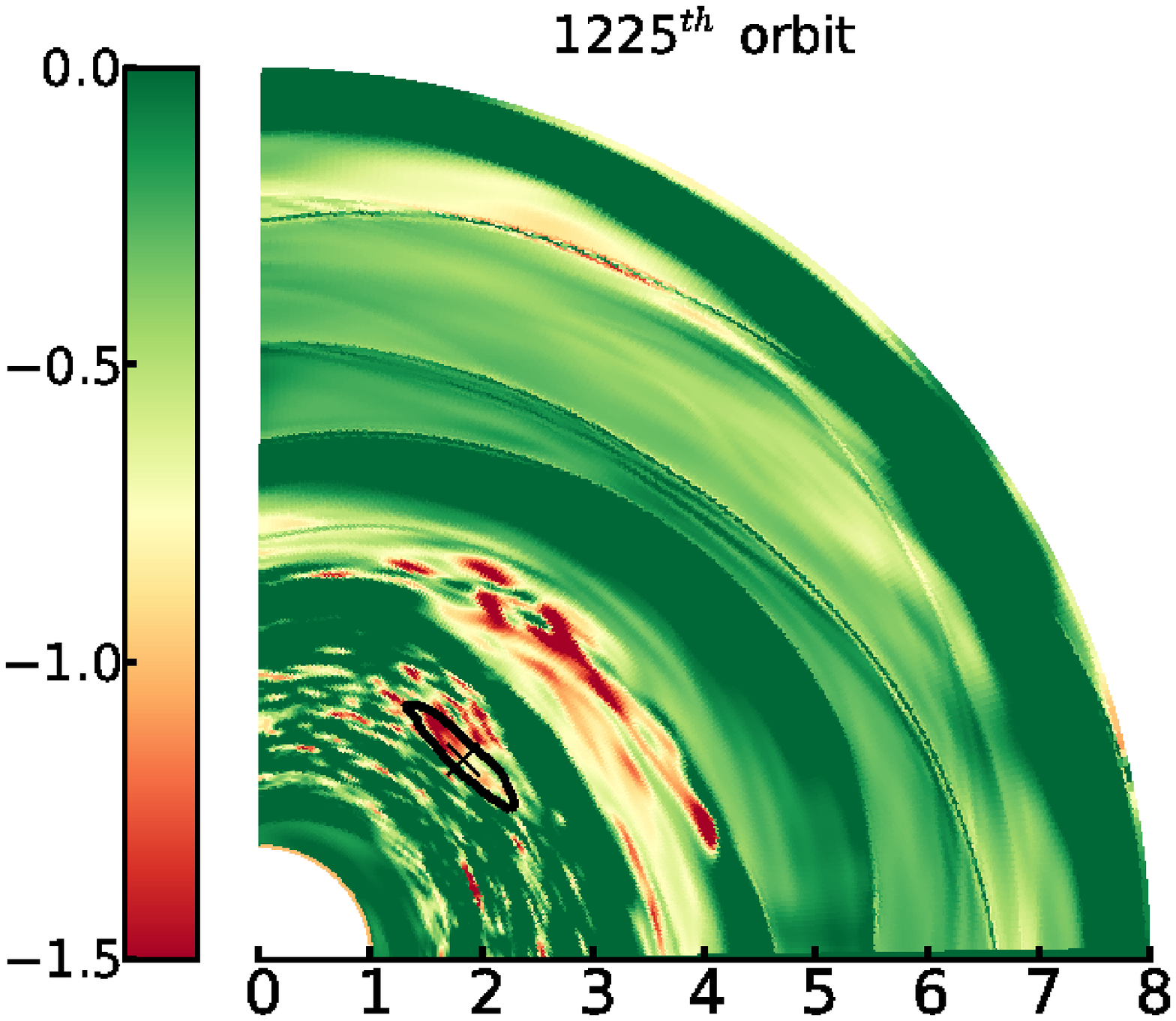}
\includegraphics[scale=0.35]{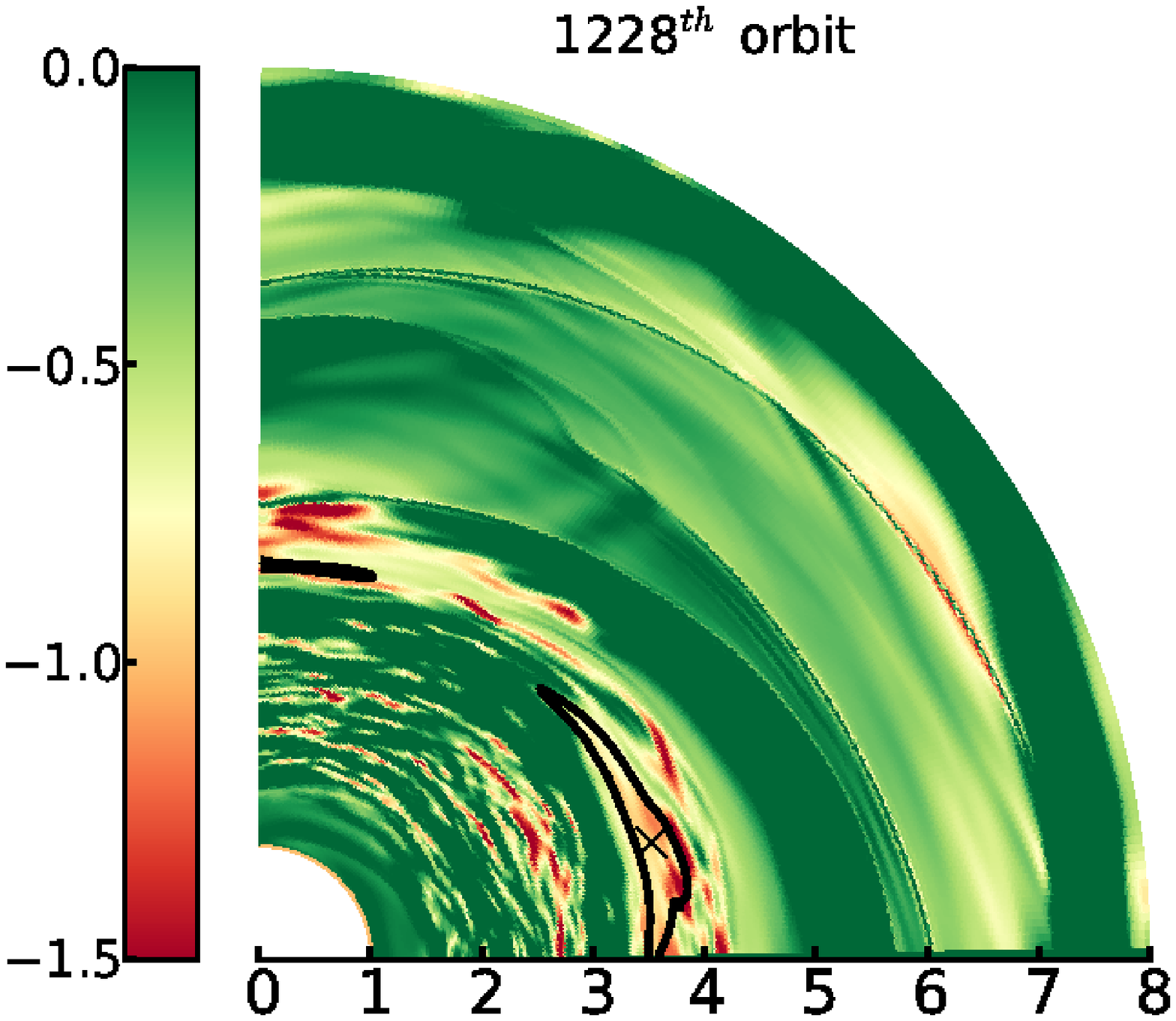}
\includegraphics[scale=0.35]{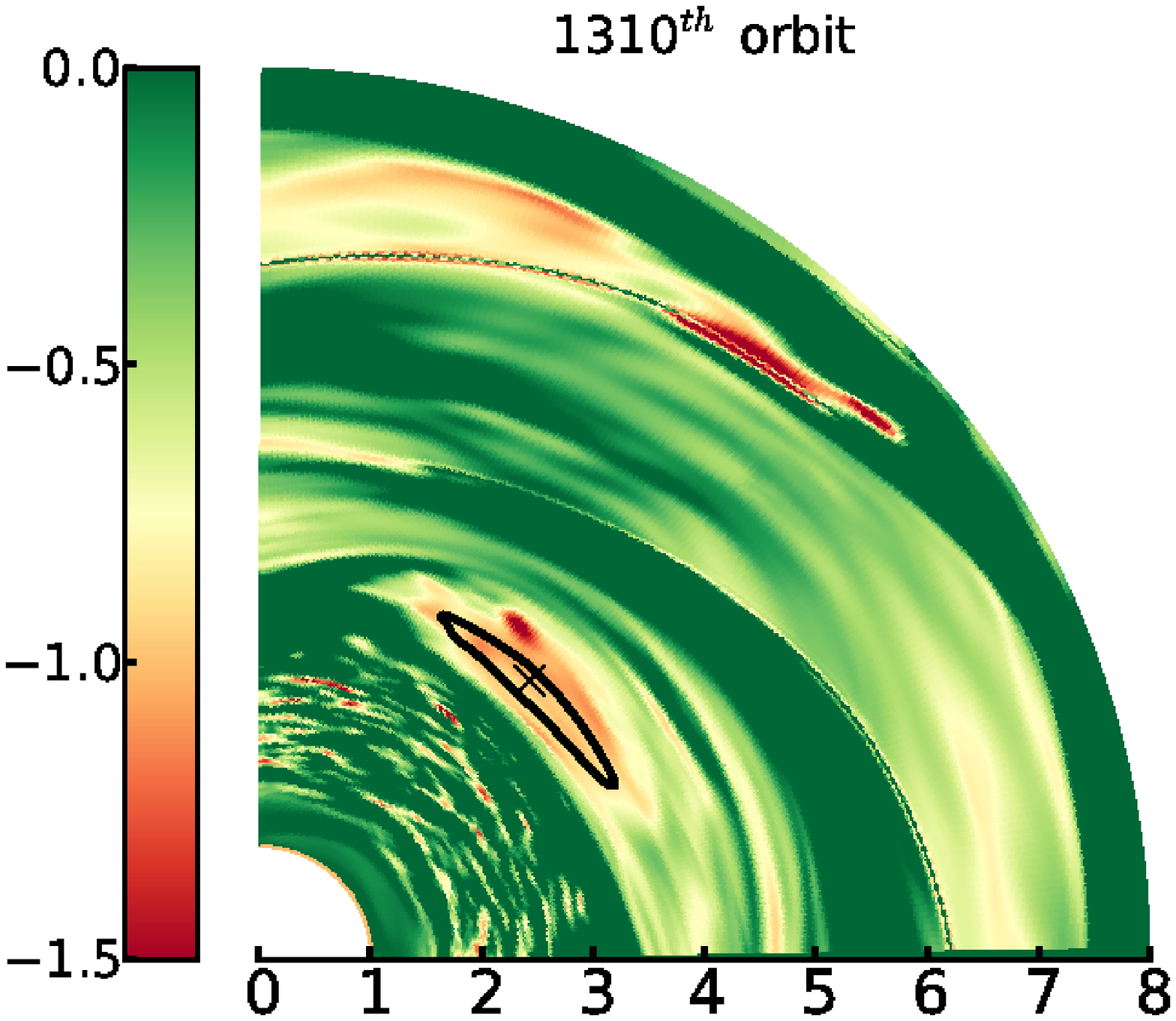}
\caption{From left to right and top to bottom: snapshot of the
  relative vorticity perturbation at the $1087^{th}$, $1124^{th}$,
  $1173^{th}$, $1208^{th}$, $1216^{th}$, $1225^{th}$, $1228^{th}$ and
  $1310^{th}$ orbit. The relative vorticity perturbation has been
  vertically averaged. We draw with a black line the iso-contour
  corresponding to
  $80\%$ of the density maximum $\rho_m$. The location of the density
  maximum is shown by the black cross.}   
\label{vortex1}
\end{center}
\end{figure*}

\begin{figure}
\begin{center}
\includegraphics[scale=0.4]{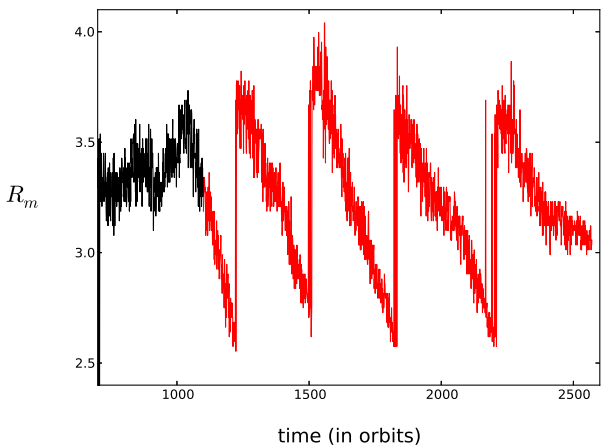}
\caption{Radial position of the density maximum in the 3D run
  during the static dead-zone phase (in black) and the self-consistent
  phase (red curve).}
\label{position3}
\end{center}
\end{figure}

\begin{figure}
\begin{center}
\includegraphics[scale=0.4]{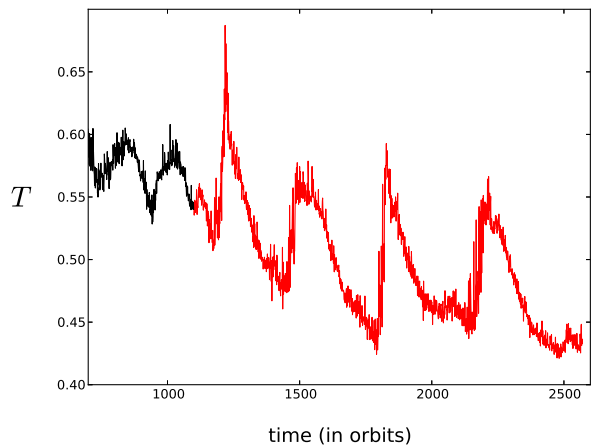}
\caption{Temperature averaged over the vortex area in the 3D run
  during the static dead zone phase (in black) and the self-consistent
  phase (red curve).} 
\label{temp3D}
\end{center}
\end{figure}

\begin{figure}
\begin{center}
\includegraphics[scale=0.4]{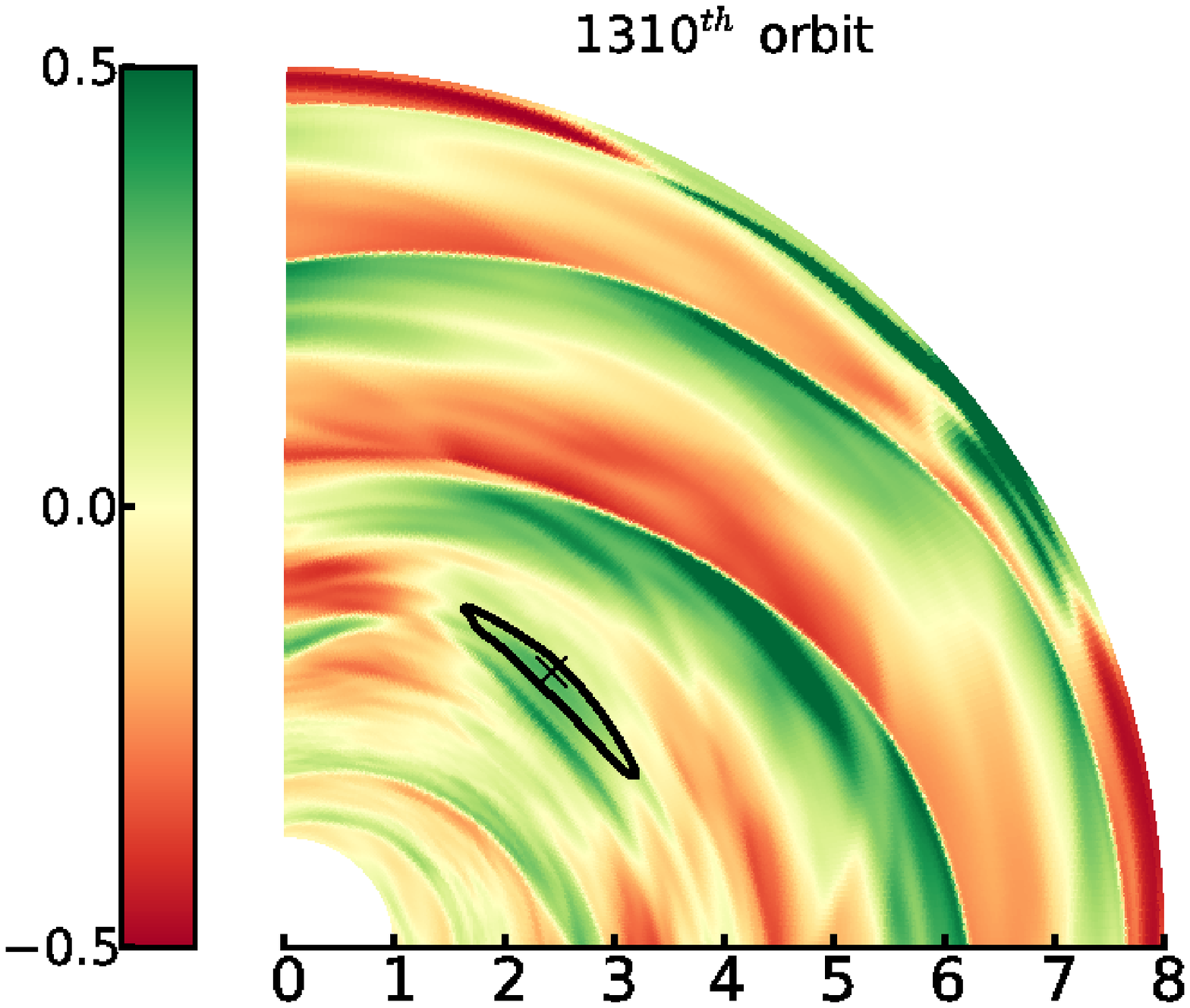}
\caption{Snapshot of the
  relative density perturbation at the
  $1310^{th}$ orbit. The relative density perturbation has been
  vertically averaged. we draw with a black line the iso-contour of
  $80\%$ of density maximum $\rho_m$. The location of the density
  maximum is shown by the black cross.}
\label{wavpatern}
\end{center}
\end{figure}

In the final phase of the simulation
 we restart the simulation at $t=1100$
inner orbits and close the feedback loop: the resistivity is now a
function of temperature according to Eq.~(\ref{resist}). 
With this configuration, \citet{MOI} have shown that
the dead zone inner edge moves radially before stalling at a
critical radius where the gas temperature is
\begin{equation}
\label{criteria}
T(R_c)=\frac{5}{4}T_\text{MRI} \, .
\end{equation}
That critical temperature is shown on figure~\ref{tempfull_turb} and gives
$R_c=3.7$. In practice, we found that the dead/active interface
initially located at $R=3.5$ remains at that position over the simulation.

We found that the vortex formed during the static dead-zone step
immediately begins to migrate
inward. Its behavior is illustrated by the sequence of successive
snapshots displayed in figure~\ref{vortex1} where we plot
the vertically integrated vertical relative vorticity $\delta
\omega_z$ of the gas 
\begin{equation}
\delta \omega_z=\frac{\del \btimes \bb{v}-\del \btimes \bb{v_K}}{|\del
  \btimes \bb{v_K}|} \bcdot \bb{e_z}  \, ,
\end{equation}
where $\bb{v_K}$ stands for the Keplerian
velocity\footnote{Using the background velocity (i.e. taking into
account the modifications to Keplerian rotation induced by
pressure) makes very little difference to $\delta \omega_z$}.
The first five panels of figure~\ref{vortex1} clearly show the
vortex's migration into the active zone. Note also that as it moves
toward the star the vortex becomes smaller and smaller until it
disappears after $150$ orbits (panel $6$). But by this stage a new vortex
has begun to form at the dead-zone inner edge (panel $7$) and shortly
begins its inward migration about ten inner orbits later (panel
$8$). Following snapshots (not shown) indicate that it will also
penetrate into the active zone before being similarly disrupted. 

In figure \ref{position3}, we plot the evolution of the vortex's
radial position $R_\text{vort}$ with time. This is calculated from
the position of the density maximum rather than the vorticity minima
because the turbulent
fluctuations in the active zone can complicate the identification of the
vortex.
Figure \ref{position3} shows that vortices follow a cycle of
formation, migration, and disruption with a period of about $200$ 
inner orbits. Vortices migrate inwards with a velocity $\sim 10^{-3} R_f
\Omega_f$ (where $R_f$ and $\Omega_f$ respectively stands for the
vortex formation radius and the angular velocity at that location)
from $R \sim 3.7$ to $R \sim 2.6$, which corresponds to about $3$
scale heights at that location.

By definition, the temperature within the vortex at the
formation location is below the MRI activation threshold (simply
because the vortex forms in the dead zone) and the flow is relatively
laminar. Heating inside the vortex is consequently low while the
cooling rate is slightly increased because of its larger density
compared to its surroundings. The combination of such reduced heating
and  increased cooling rates induces a temperature decrease within the 
vortex during its migration, as shown in
figure~\ref{temp3D}. This causes the temperature in the vortex
to remain smaller that $T_{MRI}$ during its lifetime. The typical
cooling rate is of order $4.5 \times 10^{-3} T_{\textrm{MRI}}
\Omega_0$ which results in the vortex temperature decreasing by about
$20\%$ by the time it disrupts. Note also the secular
decrease in the vortex temperature over many cycles.
This indicates that the vortices' properties are still evolving
cycle after cycle which raises the question of the existence of a stationary regime.
We will come back to this issue in section 3 by performing 2D long
term simulations.

Finally, we note that the temperature in the dead zone is strongly correlated
with the presence of the vortex. Indeed, dead-zone temperatures increase from
$T\sim 0.4$, when the vortex is weak at the end of its lifespan, to
$T\sim 0.7$ just after a new one has formed. This is due to
the presence of a pattern of strong density waves
excited by the vortex (see the map of the density perturbations
$\delta\rho/ \langle \rho \rangle$
on figure~\ref{wavpatern})\footnote{In the remainder of the paper, the
symbol $\langle . \rangle$ denotes an azimuthal, vertical average.}. The
density increase at the crest of these waves is significantly larger
($\sim 50\%$ of the local density) than described in \citet{MOI} in the absence of
a vortex (only $\sim 10\%$ of the local density). These waves provide
an additional source of heating and explain the warmer dead zone
temperatures we obtain here. 

\section{The 2D simulation}
\label{2D}

In the 3D simulation, the flow is complex and the different physics ---
turbulence, MHD, and thermodynamics --- are difficult
to disentangle. In addition, the large computational cost associated
with such simulations precludes long term integration. For example,
the temperature evolution displays a systematic drift over a few
cycles (see figure~\ref{temp3D}), thus raising the question of whether
the system is able to settle into a quasi steady state, or whether the
cycle described in the previous section is transient. In order to
answer these questions, we present a non-magnetized laminar 2D viscous
simulation which also reproduces the
vortex cycles. The simpler setup eases the interpretation of
the results and their long term relevance and highlights potential
limitations to our simulations. In this section, we briefly present the 2D
model before describing in detail the formation, migration and
disruption phases of a typical cycle in that situation. 

\subsection{Setup and run parameters}
\label{2drun_sec}
We performed a purely hydrodynamic laminar 2D simulation 
of a PP disk, similar in every other way to the 3D simulations of
Section 2.
The evolution of the disk model described below is calculated using the uniform grid version of the RAMSES code.
The equations we solve are:
\begin{eqnarray}
\frac{\partial \Sigma}{\partial t}+\del \bcdot (\Sigma \bb{v}) &=& 0 \label{cont_eq}\\
\frac{\partial \Sigma \bb{v}}{\partial t}+\del \bcdot (\Sigma \bb{v} \bb{v}) + \del P &=&
-\Sigma \del \Phi + \del \tau \label{mom_eq}\\
\frac{\partial E}{\partial t}+\del \bcdot \left[ (E+P) \bb{v}
 \right] &=& -\Sigma\bb{v}\cdot\nabla \Phi
-{\cal L} + \del \bcdot (\tau \cdot \bb{v})\label{energy_eq}
\end{eqnarray}
where $\tau$ is the Navier-Stokes stress tensor and $\Sigma$ the disk
surface density. We use the viscous prescription as a crude model of
both turbulent angular momentum transport and heating. The kinematic viscosity
radial profile is calculated using 
\begin{equation}
\nu=\langle \alpha \rangle_t c_s^2/ \Omega
\end{equation}
where $\langle \alpha \rangle_t$ is the $\alpha$ profile azimuthally,
vertically and time averaged over the last $200$ orbits of the 3D
static dead zone step. Here, $c_s$ and $\Omega$ are the local sound 
speed and the angular frequency. In the 3D static dead-zone step the
vortex launches waves that significantly increase the angular momentum
transport in the dead zone \citep[see][]{MOI}. Hence the $\langle
\alpha \rangle_t$ profile has two contributions coming from the turbulence and
from the waves. In order to remove the wave contribution in the 2D
simulations, which will emerge self-consistently from the vortex,
 we take $\langle \alpha \rangle_t=0$ outward of $R=3.5$.
In the disk active zone, $\langle \alpha \rangle_t$ steadily
  decreases from about $0.04$ at $R \sim 2$ to $0.01$ at $R=3.5$.
The viscosity jump at $R=3.5$ leads to the formation of a pressure/density
bump at the outer edge of the viscous region (or, equivalently, at the
inner edge of the inviscid/dead zone). We checked that the bump
mass-loading rate is identical to that obtained in the 3D simulations,
thus validating the $\alpha$--prescription. This corresponds to the
static dead-zone phase of Section 2.

As soon as the density
bump has reached five times the initial local density (which occurs at
$t=1000$), we add a random velocity perturbations at the position of the bump
to trigger the RWI
and then close the feedback loop between resistivity and temperature by
imposing: 
\begin{equation}
\label{visc_feed_back}
\nu(T)=
\begin{cases}
  \langle \alpha \rangle_t c_s^2/\Omega \, \, \, \,
  \textrm{if} \, \, \, T>T_\text{MRI}\\ 
  0 \, \, \, \, \, \textrm{otherwise}.
\end{cases} 
\end{equation}
The perturbation amplitude equals $10 \, \%$ of the sound speed so that 
it mimics the velocity fluctuations induced by the turbulence on the
bump. The influence of the initial bump size on the results will be
discussed in section~\ref{migration_disruption}. 
We have hence entered the analogue of the self-consistent dead-zone
phase of Section 2.
A vortex forms at $R \sim 4$ in a few tens of orbits. We concentrate
in the following section on a detailed analysis of this phase, for
which the idealized conditions of the 2D simulations provide a
favourable environment. 

\subsection{Vortex formation}
\label{forming}

\begin{figure}
\begin{center}
\includegraphics[scale=0.4]{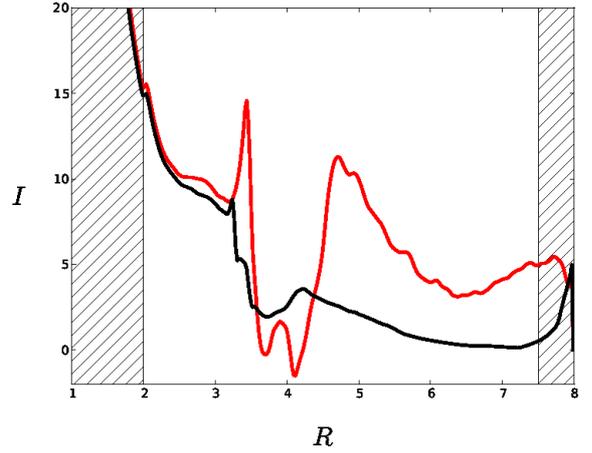}
\caption{Radial profiles of $\mathcal{I}$ at the $1000^{th}$
  orbit ({\it red curve}) and after the simulation has reached a
  quasi-steady state ({\it black curve}). $\mathcal{I}$ is averaged
  azimuthally for both cases.} 
\label{peaky}
\end{center}
\end{figure}

\begin{figure}
\begin{center}
\includegraphics[scale=0.4]{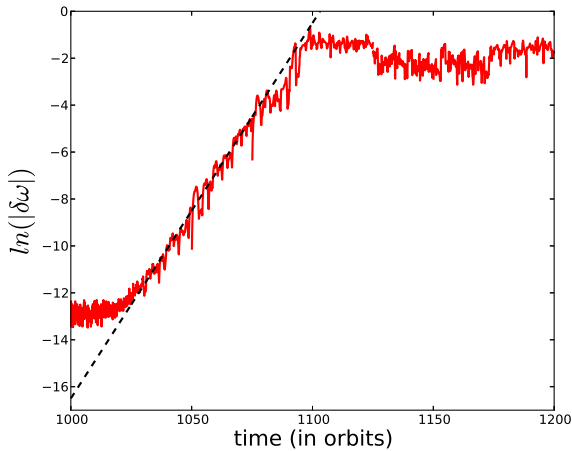}
\caption{Evolution of the logarithm of the vorticity perturbation
  $\delta \omega = \omega_{\textrm{max}}(R_\text{vort}) -  \omega_{\textrm{min}}(R_\text{vort})$ at the
    bump location in the 2D viscous simulation. The best fit of the
  linear part of the amplitude growth is plotted in black and display
  a slope equal to $\gamma/\Omega(R_f)=0.2$.}
\label{grosserate}
\end{center}
\end{figure}

The origin of the vortex forming at the dead/active interface is most
likely due to the RWI \citep{lyra&maclow12,
  2012MNRAS.419.1701R}. A 2D adiabatic disk is unstable to the RWI where
\begin{equation}
\label{peaky_eq}
\mathcal{I}=\frac{\Sigma\, (\nabla\times\mathbf{v})\cdot \mathbf{e_z}}{P^{\frac{2}{\gamma}}},
\end{equation} 
a conserved quantity similar to the
potential vorticity, possesses a local extremum
\citep{1999ApJ...513..805L}. Here $\Sigma$ is the equilibrium surface
density and $\gamma$ is the adiabatic index.
We show on figure~\ref{peaky} the profile of $\mathcal{I}$ at $t=1000$
(i.e. when random velocity fluctuations are added to the
flow).
Aside from small-scale variations, $\mathcal{I}$ possesses a
deep and long-lasting trough at the vortex location, strong
evidence that the
RWI is active and generates the vortex. Note that our simulations are diabatic, and
hence do not strictly conserve $\mathcal{I}$, but that this fact does
not impinge on (and in fact probably aids) instability here.
In the 3D simulation, the vortex and the bump form simultaneously
which complicates a clear identification of the RWI's linear phase. 
Nonetheless,
the profile of $\mathcal{I}$ at the beginning of every cycle
displays a similar trough at the position where the next vortex will
form (not shown). Taken together, the results of the 2D and 3D simulations
consistently point toward the RWI as being responsible for the vortex
growth in our simulations.

The growth of vorticity perturbations are presented in
figure~\ref{grosserate}.
The perturbation grows exponentially with a growth rate of about
$\gamma/\Omega(R_f) \sim 0.2$. For the density bump properties (height and
width), the empirical law of
\citet{2013MNRAS.430.1988M} gives a growth rate about twice as large, such
that $\gamma_{th}/\Omega(R_f) \sim 0.5$. Note however that this relation was
obtained for globally isothermal disks with a small density
perturbation such as $\delta\rho/\rho<0.3$. In our 2D simulation the
density bump is $10$ times larger and the temperature gradient at the
dead zone inner edge is significant. The empirical
relation still provides a useful rough estimate and sanity check.

When the period of the circular motion around the vortex centre becomes comparable to the growth time
scale (at $t \sim 1100$) the growth ends. The vorticity
  perturbation is then $\delta \omega \simeq -0.05 \Omega_0 \simeq 0.4 \Omega(R_f)$, 
corresponding to twice the growth rate, as noticed by
\citet{2013MNRAS.430.1988M} in their simulations.
By this stage the vortex has already begun to migrate while capturing
an important fraction of the initial bump's mass: 
\begin{equation}
\frac{\delta \rho}{\langle \rho(Rm)
  \rangle}=\frac{\rho_{\textrm{max}}(R_\text{vort})-
  \rho_{\textrm{min}}(R_\text{vort})}{\langle \rho(Rm) \rangle} \sim 0.4 \,,
\end{equation}
thus leaving about $60 \%$ of the bump mass behind at $R=4$. 

\subsection{Vortex migration and disruption}
\label{migration_disruption}

\begin{figure}
\begin{center}
\includegraphics[scale=0.4]{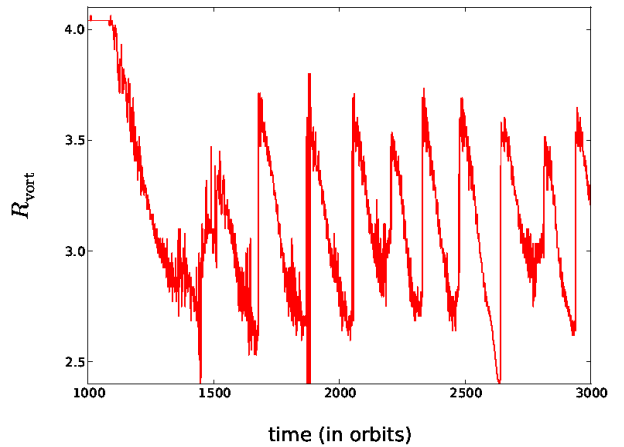}
\caption{Radial position of the density maximum in the 2D viscous
  simulation using the RAMSES code.}   
\label{position2}
\end{center}
\end{figure}

\begin{figure*}
\begin{center}
\includegraphics[scale=0.3]{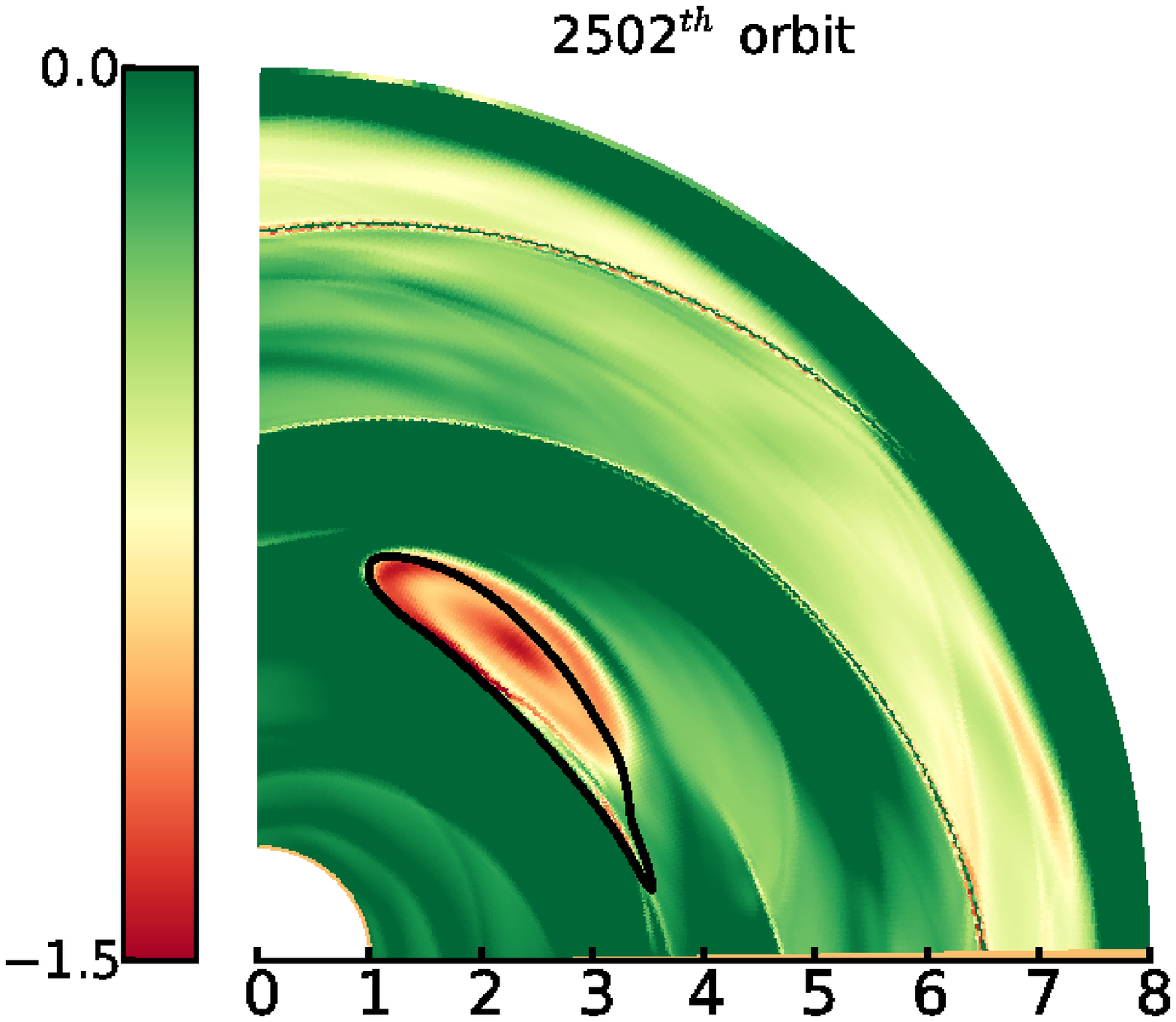}
\includegraphics[scale=0.3]{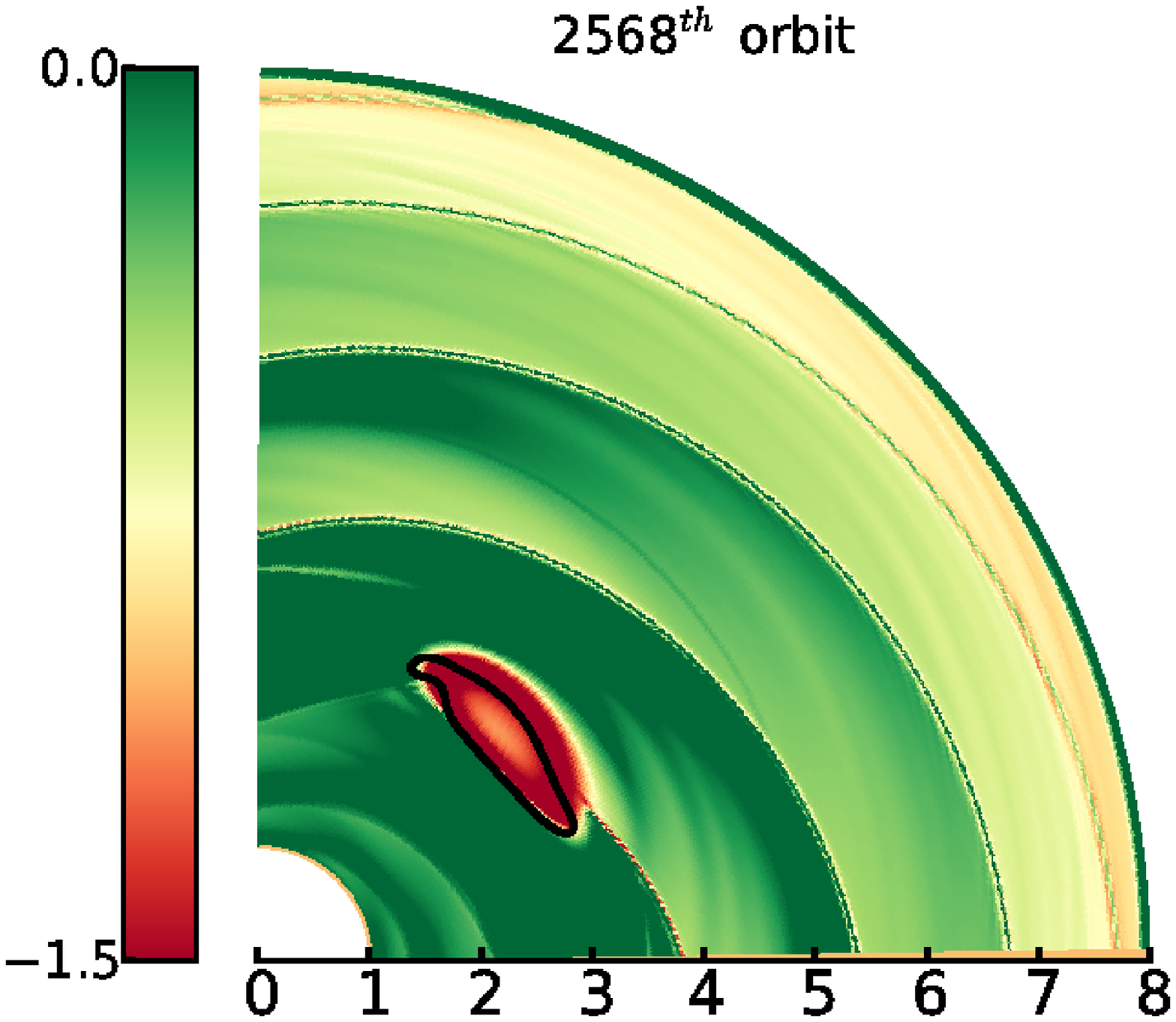}
\includegraphics[scale=0.3]{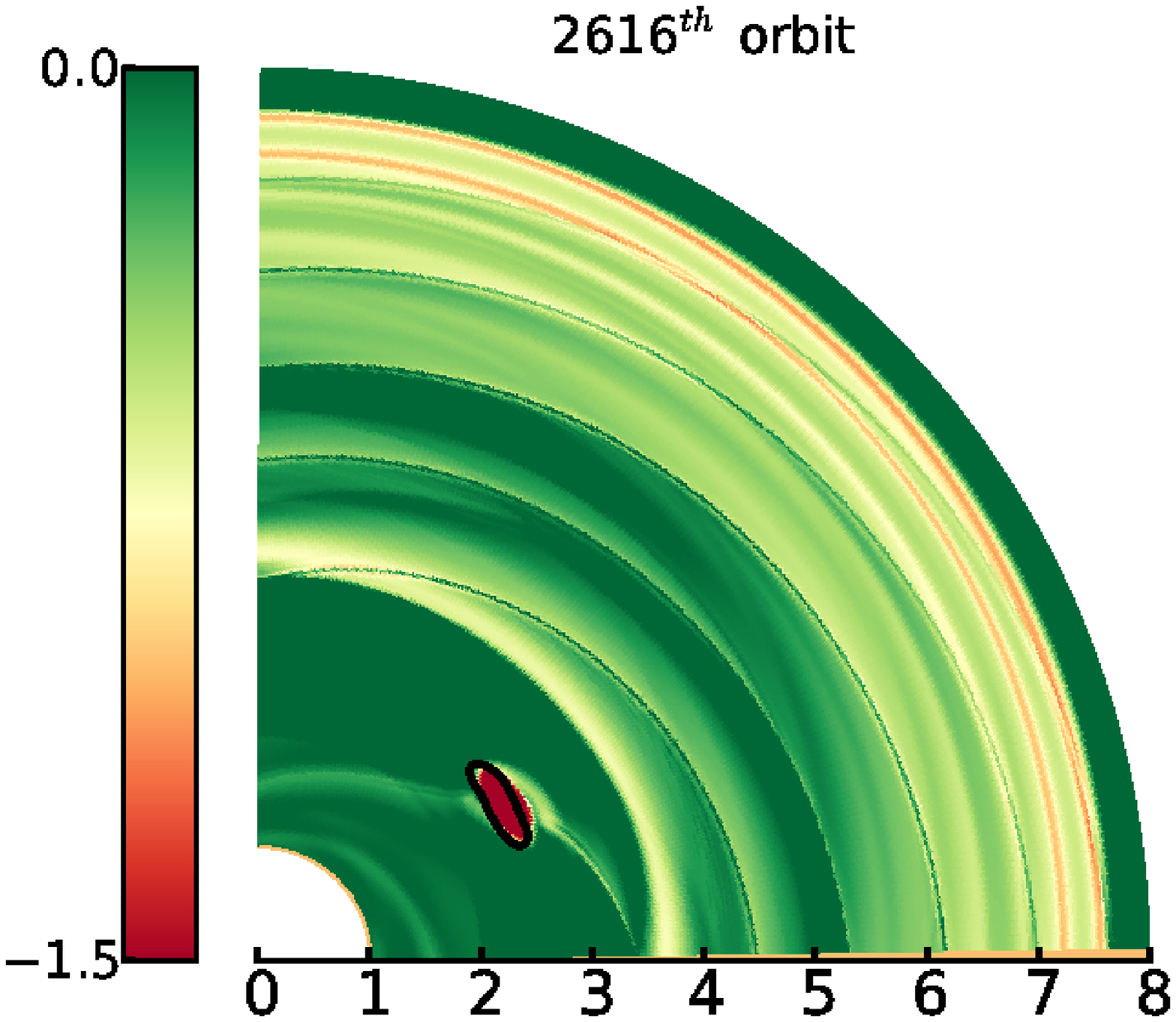}
\includegraphics[scale=0.3]{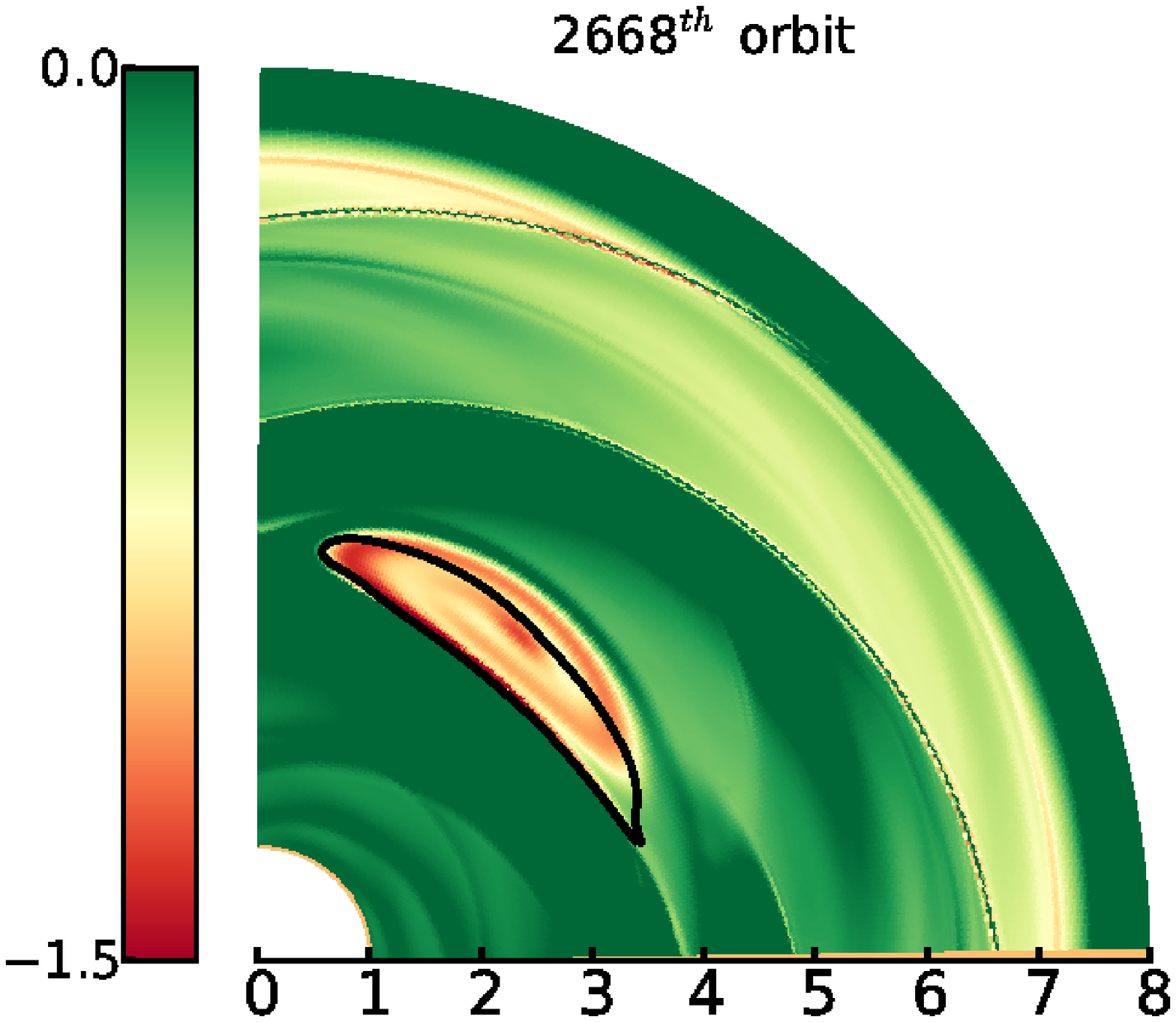}
\caption{From left to right and top to bottom: snapshot of the relative vorticity
  perturbation at the $t=2502$, $2568$, $2626$ and
  $2668$. The relative vorticity perturbation has been vertically
  averaged. We draw with a black line the iso-contour of $80\%$ of
  density maximum $\rho_m$. The location of the density maximum is
  shown by the black cross.}
\label{vortex2}
\end{center}
\end{figure*}

\begin{figure}
\begin{center}
\includegraphics[scale=0.4]{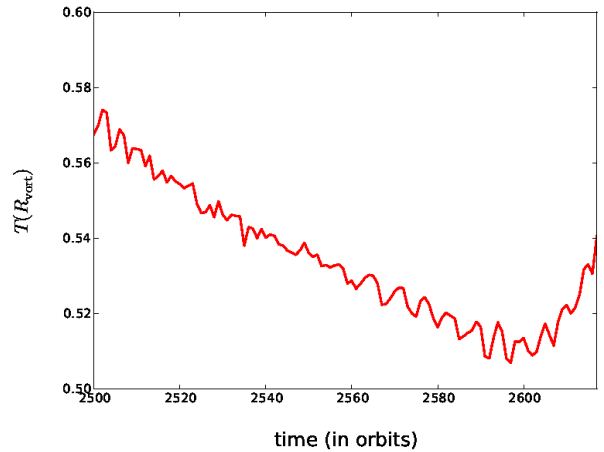}
\caption{Evolution of the mean temperature inside the vortex during the cycle starting at the $t=2500^{th}$ inner orbit.}
\label{tempi}
\end{center}
\end{figure}

\begin{figure}
\begin{center}
\includegraphics[scale=0.4]{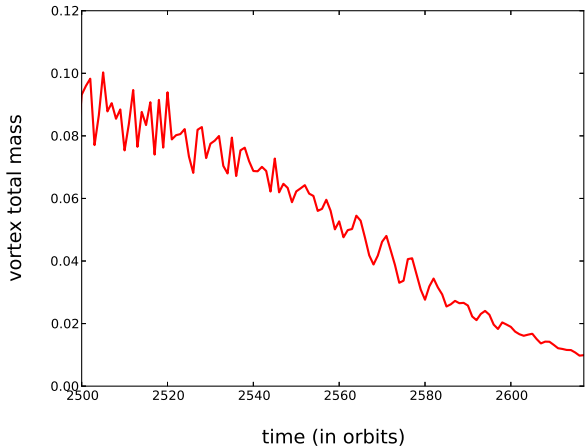}
\caption{Evolution of the vortex total mass over the cycle starting at 
  $t=2500$, given in unit of the initial disk mass. It is found to
  decrease steadily during the vortex migration.} 
\label{rho}
\end{center}
\end{figure}

After the first appearance of a vortex, we
observe during the remainder of the 2D simulation a vortex life cycle 
similar to that obtained in the 3D simulation. As in the 3D
simulation, the cycle properties change over the first few cycles,
before the disk relaxes into a well-defined quasi-periodic state (see
figure~\ref{position2}). During that quasi-periodic evolution,
the profile of $\mathcal{I}$ at the beginning of a cycle has converged
to the black curve of figure~\ref{peaky}. The existence of a minimum
of $\mathcal{I}$ at a well defined radius during the remainder of the
the disk evolution shows that the conditions for vortex formation
persist to late times, presumably driven by the continuing accretion
mismatch at this point. The important difference is that the profile of the black
curve is self--consistently obtained as a result of the disk
evolution. Additional simulations starting with different bump sizes   
were found to exhibit different relaxation phases but all converged to
the same quasi-periodic state. These results, obtained using the 2D
simulation, strongly suggest that the 3D simulation of section 2
would also evolve toward a quasi stationary state if evolved for
longer. Indeed, during this relaxation phase, the vortex migration 
range is similar to that observed in the 3D simulation. Forming at $R
\sim 3.7$, vortices migrate inwards to $R \sim 2.7$. However, such a
similarity does not extend to all the simulation diagnostics. For
example, the cycle period is two times shorter and vortices migrate
twice as fast than vortices in the 3D run. We will come back to this
issue in the discussion. 

We now investigate the properties of the quasi-periodic phase by focusing on
one cycle, starting at $t=2500$. At that time, a vortex
has formed and is about to start migrating inwards. Its relative density
perturbation at this point is $\sim 0.4$.
Figure~\ref{vortex2} presents a series of snapshots showing the vorticity in
the disk at four different times. The first panel
illustrates the state of the disk at the beginning of the vortex: it
is clear that a vortex exists at $R=3.7$. In panels 2 and 3
($50$ and $100$ orbits later) the vortex has drifted closer to
the star and has shrunk as it does so. Finally, the last
panel shows the vortex has disappeared but
a new vortex has formed at a larger radii at the interface. Overall, the
evolution is very similar to that obtained in the 3D simulation (see
figure~\ref{vortex1}). As shown in figure~\ref{tempi}, the
similarity extends to the temperature evolution of the vortex: as in
the 3D simulation, the vortex cools while it moves to hotter
region. The cooling rate during the stationary cycles is found to be
 $\approx 6\times 10^{-3} T_{\textrm{MRI}} \Omega_0$, i.e. slightly larger
than obtained in the 3D simulations. Finally, we show in
figure~\ref{rho} the evolution of the mass trapped inside 
the vortex. As suggested by the snapshots shown on
figure~\ref{vortex2}, the vortex loses mass during the cycle. 

\subsection{A numerical test}
\label{pluto_test_sec}

\begin{figure}
\begin{center}
\includegraphics[scale=0.4]{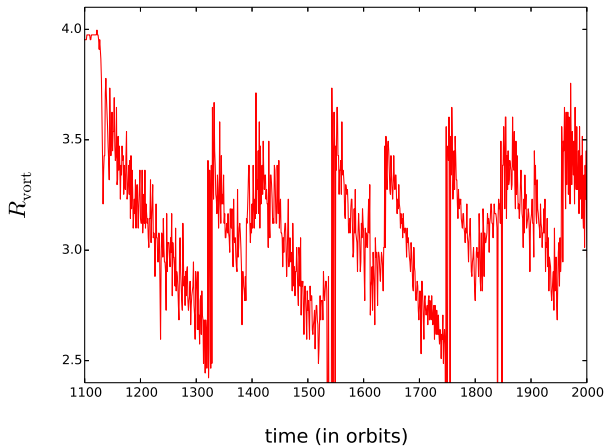}
\caption{Radial position of the density maximum in the 2D viscous
  simulation using the PLUTO code.}   
\label{position2PLUTO}
\end{center}
\end{figure}

In order to check the reliability of our simulations, we
have reproduced the simulation presented in this section using the
PLUTO code \citep{2007ApJS..170..228M}. For this run, we used a
simplified function for $\langle \alpha \rangle_t$: 
\begin{equation}
\langle \alpha \rangle_t=
\begin{cases}
  \alpha_t \left(\frac{R}{R_0}\right)^{-1.5} \, \, \, \,
  \textrm{if} \, \, \, R < R_1\\ 
  \alpha_e \left[1-\tanh \left(\frac{R-3.2 R_0}{1.5
    R_0}\right)\right]\left(\frac{R}{R_0}\right)^{-1.5}\,  \textrm{if} \, \, \,
  R_1 < R < R_2 \\  
  0 \, \, \, \, \, \, \, \, \, \, \, \, \, \, \textrm{if} \, \, \, R > R_2
\label{alphaPluto}
\end{cases} 
\end{equation}
where $R_1=2.65 R_0$, $R_2=3.44 R_0$, $\alpha_t=7 \times 10^{-2}$ and
$\alpha_e=3.5 \times 10^{-3}$ but otherwise solve the same set of
equations~(\ref{cont_eq}), (\ref{mom_eq}) and (\ref{energy_eq}).
Figure~\ref{position2PLUTO} shows the position of the vortex that
forms in this simulation. The vortex cycle is clearly reproduced with
a similar vortex migration timescale and amplitude as obtained with
RAMSES. The quantitative differences between the PLUTO and the RAMSES
simulations probably arise because of the adhoc radial profile of $\langle
\alpha \rangle_t$ we used in PLUTO that is slightly different than 
used in RAMSES. The difference is, however, small and the similarity
between the two simulations strengthens our main result: a vortex
cycle emerges from the interplay between dynamics and
thermodynamics at the dead zone inner edge.

\section{Discussion}
\label{disc}

In this section we use the results of the previous two sections
to illuminate the physical mechanism responsible for the vortex
cycle. 

\subsection{Vortex destruction}

\begin{figure*}
\begin{center}
\includegraphics[scale=0.25]{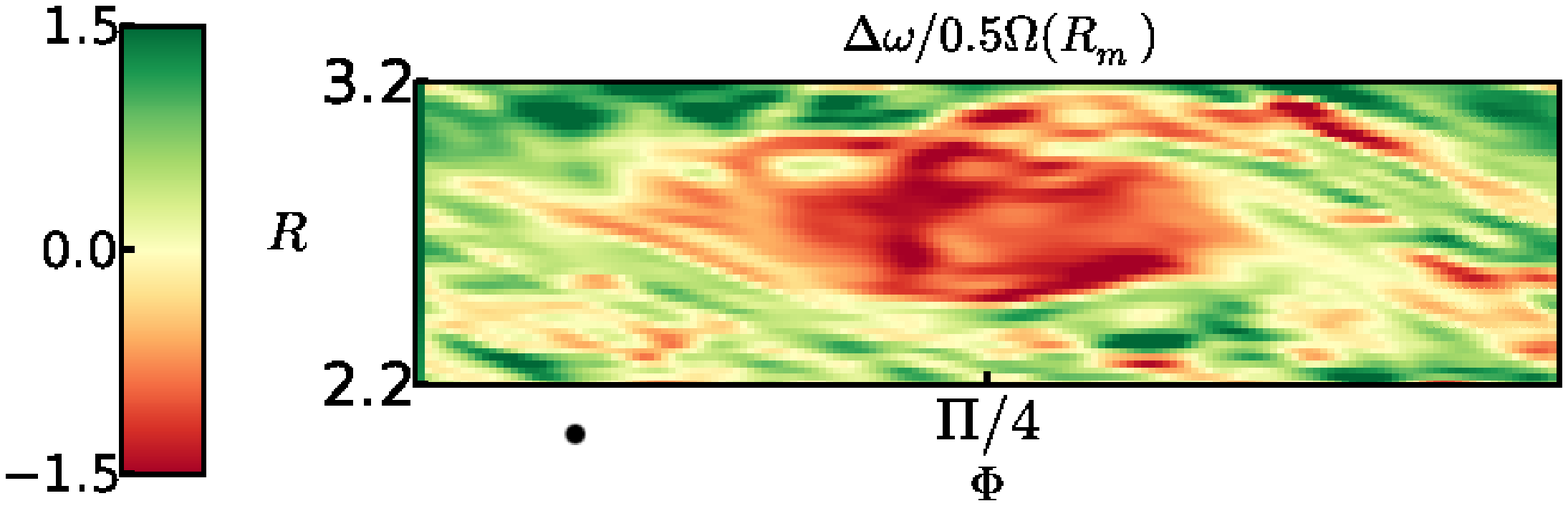}
\includegraphics[scale=0.25]{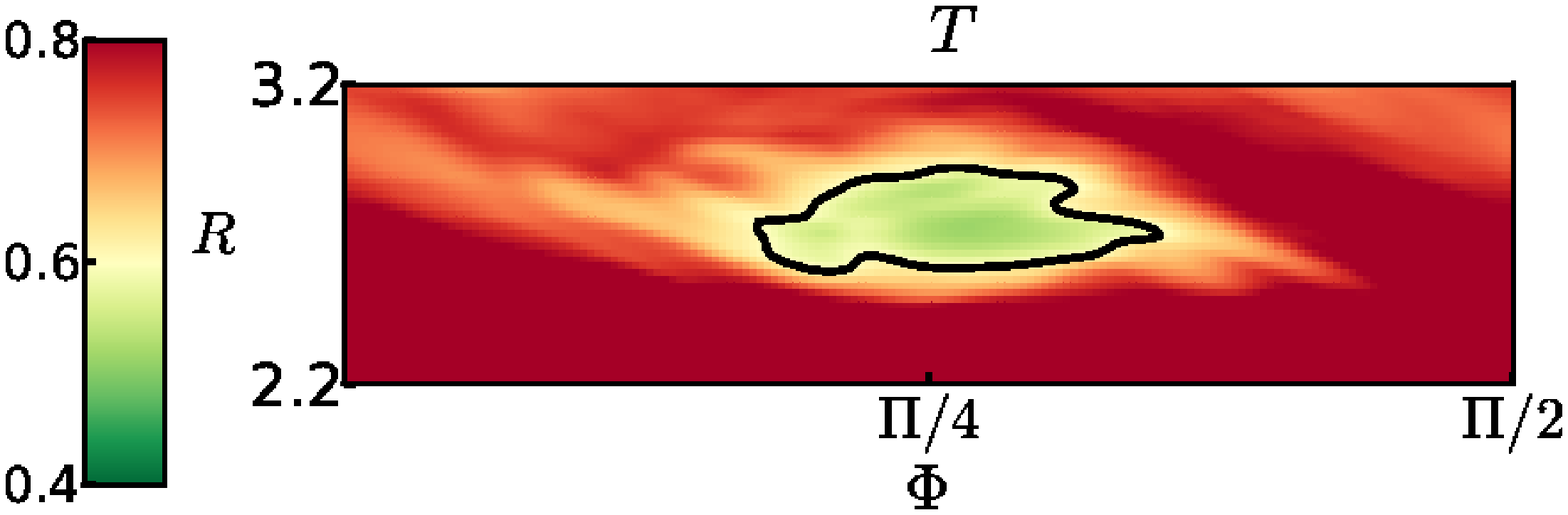}
\includegraphics[scale=0.25]{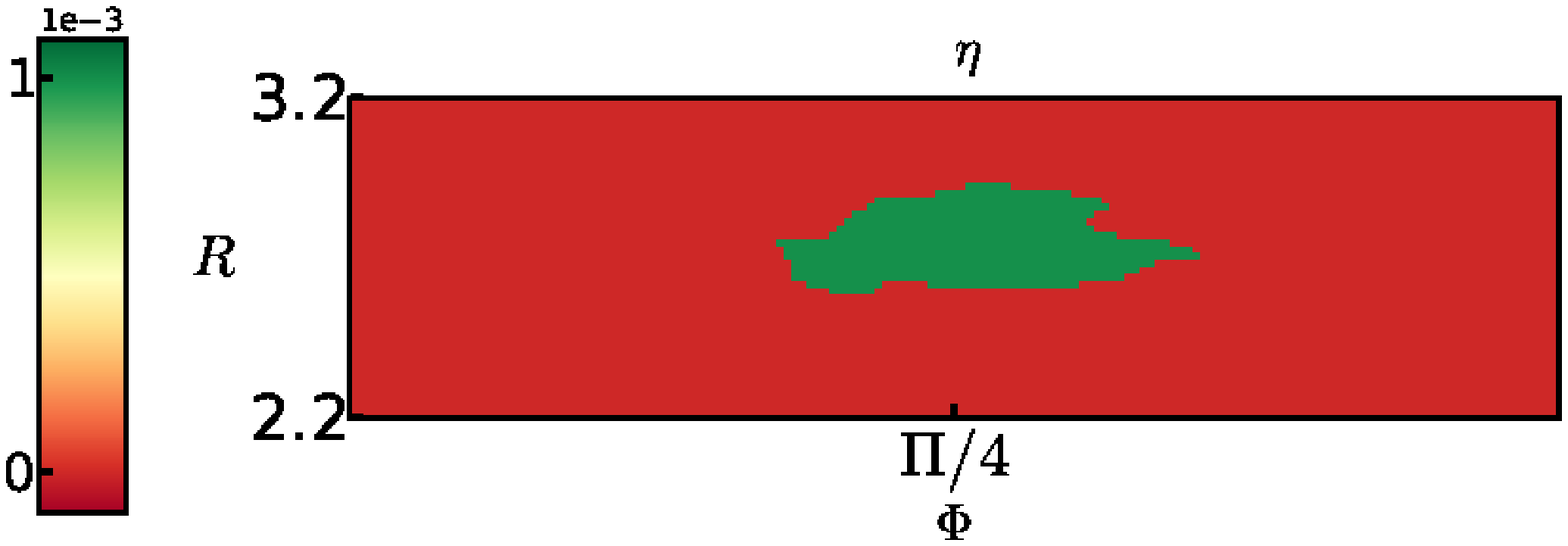}
\includegraphics[scale=0.25]{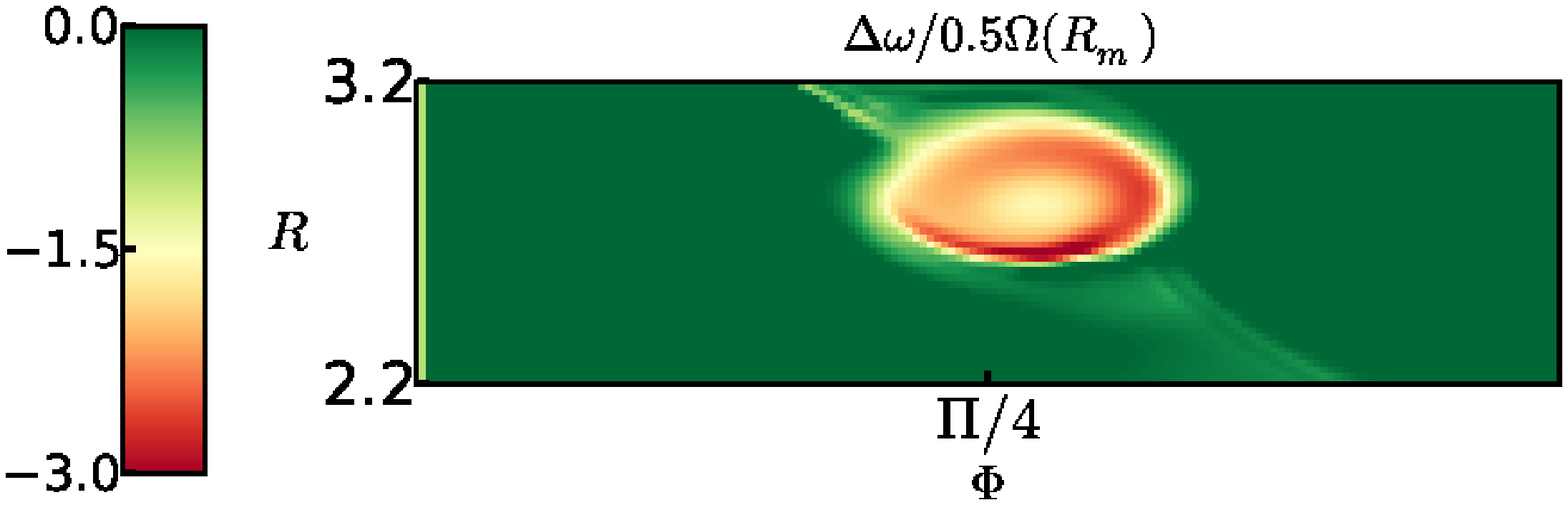}
\includegraphics[scale=0.25]{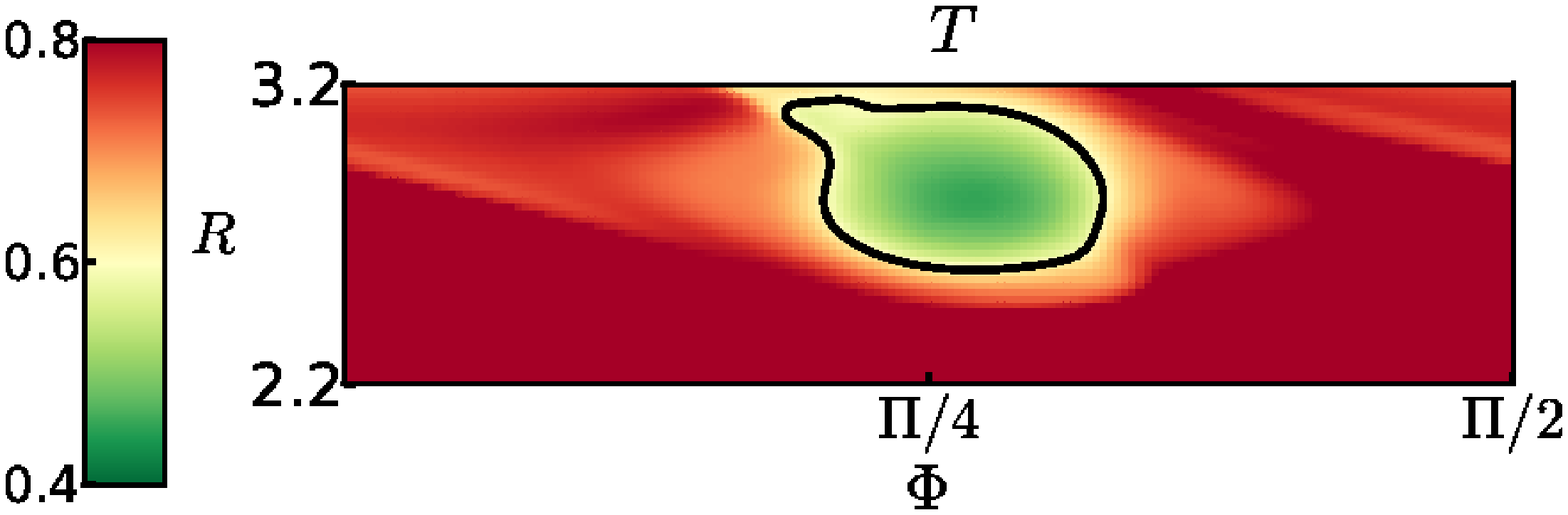}
\includegraphics[scale=0.25]{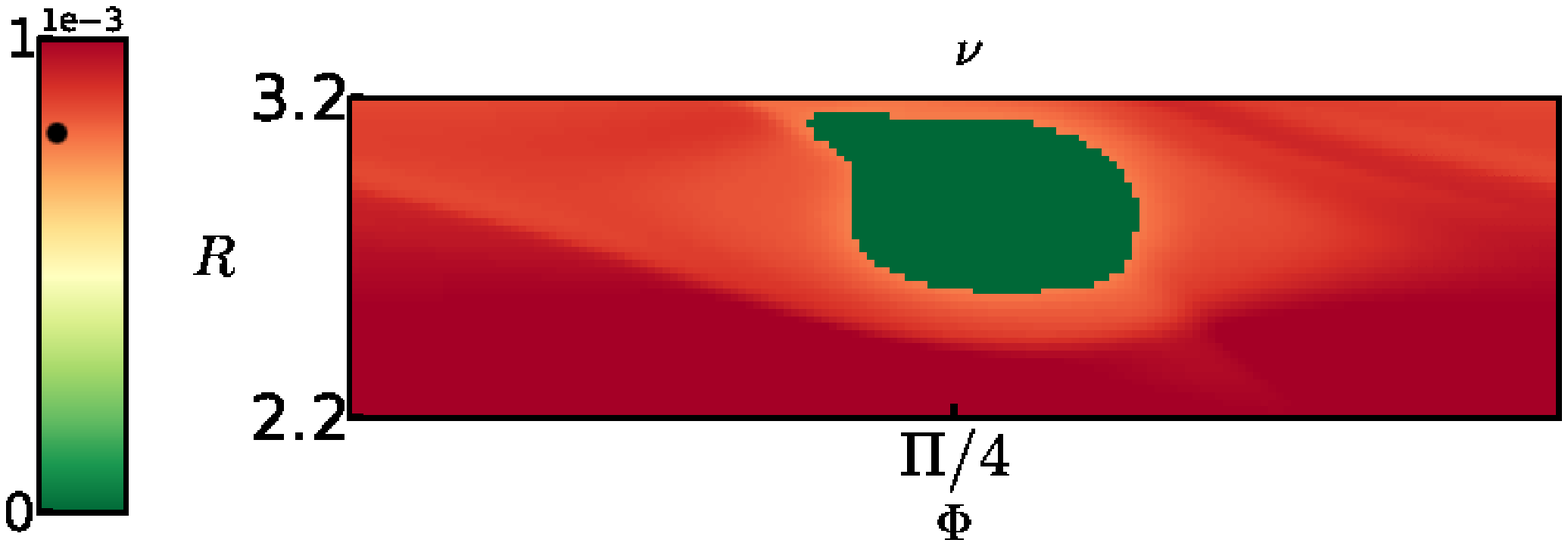}
\caption{Top panels, from left to right: snapshots of the relative
  vorticity perturbation, the temperature and the resistivity in the
  vicinity of the vortex at the $1208^{th}$ orbit, in the 3D
  simulation. The fields mapped on top panels have been vertically
  averaged. Bottom panels, from left to right: snapshots of the
  relative vorticity perturbation, the temperature and the viscosity
  in the vicinity of the vortex at the $2597^{th}$ orbit, in the 2D
  simulation. We draw with a black line the iso-contour of temperature
  $T=T_{\textrm{MRI}}$.}
\label{bandeau}
\end{center}
\end{figure*}

First we address the issue of vortex disruption, which is a little
more straightforward to understand. In
both the 2D hydro and 3D MHD simulations, the vortex forms in the
dead zone (modelled either as a highly resistive zone or as an inviscid
region) before migrating into the active region (turbulent in the 3D
case and viscous in the 2D simulation) where it gradually
disrupts. Throughout its migration the interior of the vortex is
either laminar or inviscid and therefore encounters turbulent interference at
its outer surface. In both cases this is because the temperature in
the core is below the MRI activation threshold $T_{\textrm{MRI}}$, and
hence turbulence/viscosity switches off. Because of this reason
vortices survive for relatively long times:
in a sense, they are cool bubbles of dead zone moving inside the
hot active region. 

As is clear, however, from figures 2 and 8, vortices shrink in size as
they migrate. Quantitative measurements using the 2D simulations
showed that the vortex size changed from about $0.6H$ to about
$0.25H$, at which point they dissipate. This can be understood by
realizing that vortices are overwhelmed by MRI turbulence or viscous
diffusion once their radius fall below a certain size. An estimate on the
critical size can be obtained by equating the turbulent/viscous speed
external to the vortex ($\sim \nu/s$) with the typical vortex
circulation speed, estimated from simulations ($\gtrsim s\,\Omega$).   
We find the critical vortex size $s_{\textrm{crit}}$ is given by
\begin{equation}
s_{\textrm{crit}} \sim \sqrt{\langle \alpha \rangle_t} H \sim 0.2 \, H \, .
\label{vortex_crit_size}
\end{equation}
This rough estimate is broadly in agreement with the results of the
simulation. We hence conclude that once a vortex shrinks by about
three times its original size it diffuses away in the active zone. 

An estimate on the vortex lifetime is tied to the speed at which the
vortex evolves, in particular to the speed at which it loses mass and
shrinks. The shrinking of the vortex is possibly due to the gradual
breakdown of the vortical flow via turbulent/viscous diffusion at the 
vortex surface. Here there appears a strong shear layer (see first panel in
figure~\ref{bandeau}). Thus the vortex is destroyed gradually from the outside in. And as the
outer layers disintegrate they release their mass into the surrounding
active medium.
A lower limit for the vortex lifetime is thus provided by the
diffusion timescale of vorticity over the vortex size. Using the $\alpha$ prescription, it
can be written as 
\begin{equation}
\frac{T_d}{T_{orb}}=\frac{s^2}{\nu T_{orb}}=
\frac{1}{2 \pi \langle \alpha \rangle_t}\left(\frac{s}{H}\right)^2\left(\frac{R_f}{R_0}\right)^{1.5} 
\, .
\label{visc_timescale}
\end{equation}
For the parameters of the vortex we obtained in both the 2D and 3D
simulations ($s \sim 0.5 H$, $R_f \sim 4$), this gives an estimated
destruction timescale of about $30$ orbits. Of course, this is much
shorter than the typical lifetime of about $300$ orbits we obtained in
the 3D simulations (see figure~\ref{position3}) or $150$ orbits found
in the 2D simulation (see figure~\ref{rho}), owing to the fact that
vorticity diffusion only occurs over a thin layer of size $\Delta s$
at the surface of 
the vortex. As a result, the viscosity over the vortex section is
reduced by a factor $f \sim 2 \Delta s / s$. The thickness of the
layer $\Delta s$ is very difficult to estimate. As illustrated by the
snapshots of figure~\ref{vortex2}, it is probably different in the 2D
and 3D simulations. It might also be affected by numerical
diffusion in our simulations. Nevertheless, its smallness
significantly increases the vortex lifetime compared to the estimate
given above, and might also account, at least partly, for the different
vortex lifetimes we find in the 2D and 3D simulations.

\subsection{Vortex migration}

\begin{table}[t]\begin{center}\begin{tabular}{@{}cccc}\hline\hline
Model & Cooling function & p value & Migration rate \\
& & & (in $R_f \Omega_f$) \\
\hline\hline
\textrm{ST2D} & Locally isothermal & - & $\sim 0$  \\
\hline
\textrm{RBD} & Isentropic &  $-2.75$ & $-7.8 \times 10^{-5}$  \\
\hline
\textrm{PLP1} & Isothermal & $-4$ & $-1.5 \times 10^{-3}$ \\
\textrm{PLP2} & Isothermal & $-1.5$ & $-8 \times 10^{-4}$  \\
\textrm{PLP3} & Isothermal & $0$ & $\sim 0$  \\
\hline
\textrm{PLPCOOL} & $\cal L$ (see Eq.~\ref{eq:PLPCOOL}) & $0$ & $\sim 0$  \\
\textrm{PLPCOOL+} & $\cal L_+$ (see Eq.~\ref{VortCoolAsym}) & $0$ & $-1.2 \times 10^{-3}$  \\
\textrm{PLPCOOL-} & $\cal L_-$ (see Eq.~\ref{VortCoolAsym}) & $0$ & $+1.2 \times 10^{-3}$  \\

\hline\hline
\end{tabular}
\caption{Numerical experiments reproducing the setup of
  \citet{2013A&A...559A..30R} and \citet{2010ApJ...725..146P} with
  various surface density profiles and
  thermodynamics. Simulations are 2D in polar coordinates and use a 
  resolution $(N_r,N_{\phi})=(768,1600)$.}
\label{runProperties_tab}
\end{center}
\end{table}

\begin{figure}
\begin{center}
\includegraphics[scale=0.35]{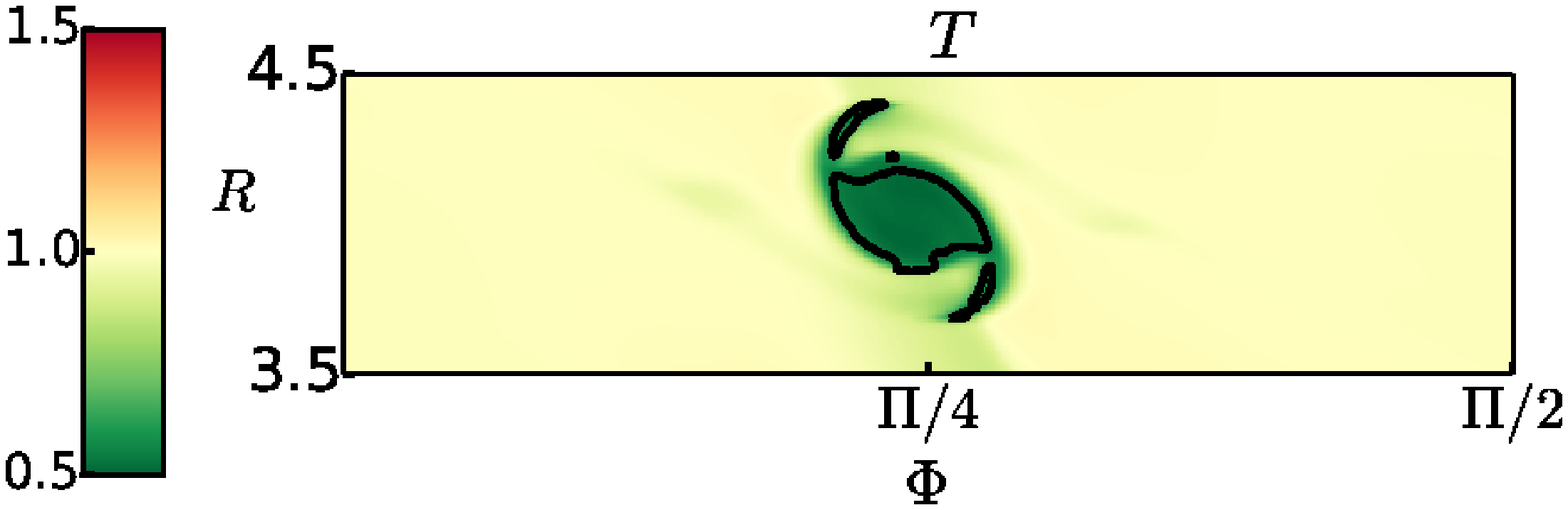}
\includegraphics[scale=0.35]{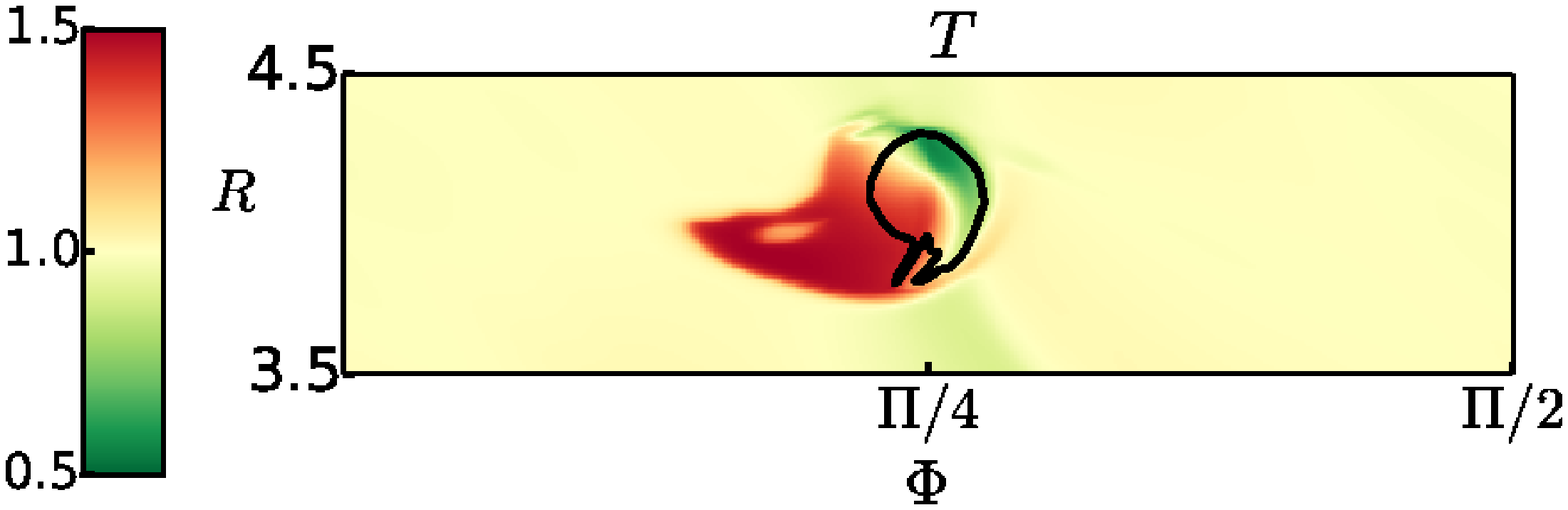}
\includegraphics[scale=0.35]{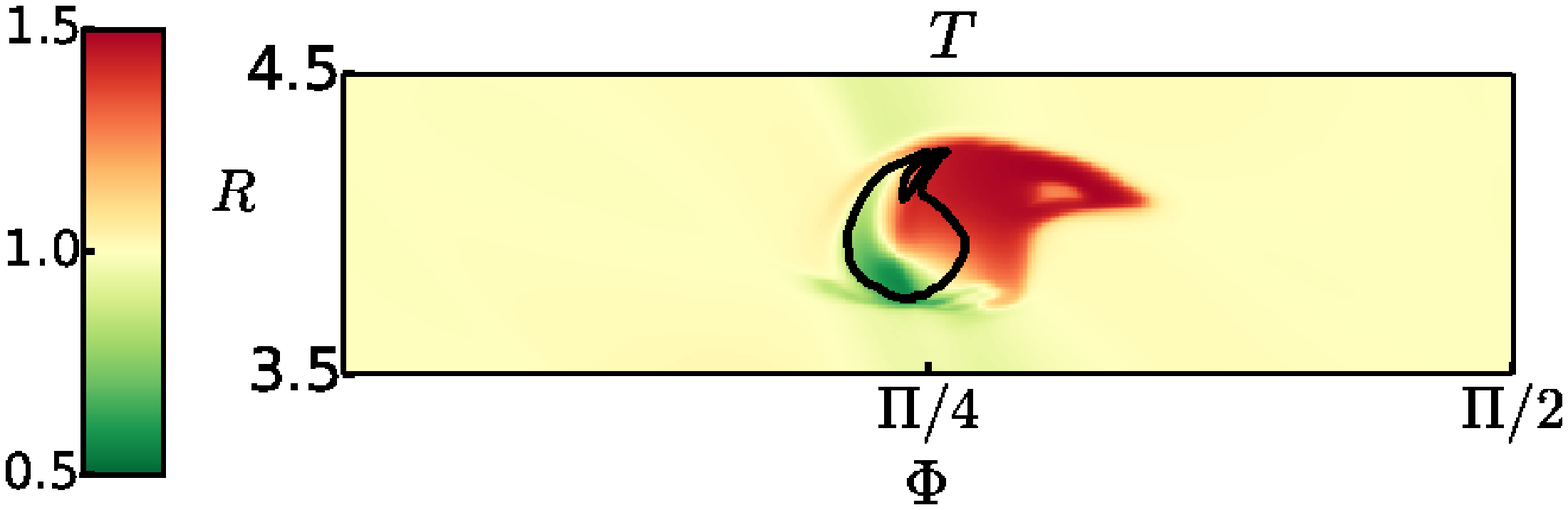}
\caption{Snapshots of the temperature (normalized by the mean temperature) in the
  vortex in model PLPCOOL ({\it top panel}), PLPCOOL- ({\it middle
  panel}) and PLPCOOL+ ({\it bottom panel}). The black line shows an
  iso--vorticity contour where the vorticy is equal to the opposite of
  the vorticity of the background Keplerian shear profile.} 
\label{bandeau_PLPCOOL}
\end{center}
\end{figure}

We now address the question of the vortex migration process. The
simple fact that we find any migration at all in our simulations is
surprising by itself. Indeed, all vortices starts their journey upon a
surface density maximum, which according to the isothermal simulations
in \citet{2010ApJ...725..146P} should fix the vortex in place.
Generally, isothermal vortices migrate toward high pressure
regions, because of asymmetric density wave launching.
 This result appears to be confirmed
by \citet{lyra&maclow12}, who report no migration of vortices in their
locally isothermal MRI-turbulent simulations of the inner dead-zone
interface, and also by 
 \citet{2012A&A...542A...9M}, who
considered the long-term evolution of a RWI vortex in
barotropic isentropic disks. However, recently
\citet{2013A&A...559A..30R}
reported inward migration in their adiabatic runs of RWI vortex
formation. Here, however, the vortex, by absorbing all the bump material, destroys
the adverse pressure gradient that would prevent its migration. In our
runs, the vortical perturbation is 
similar but the bump size at the beginning of a cycle is significantly
bigger and the vortices start their migration before they
completely absorb the bump mass. As sanity
checks, we have successfully reproduced the results of
\citet{2013A&A...559A..30R}
 albeit in a 2D simulation (see model $\textrm{RBD}$ in
table~\ref{runProperties_tab}).
Finally, the migration rate we
measure is about $10$ times higher than 
in \citet{2013A&A...559A..30R}. Questions therefore remain:
why are vortices migrating in our simulations, and why do they migrate
so fast?

We focus on the 2D simulations,
as the vertical dimension and MHD turbulence are likely to
complicate the picture. Several additional numerical experiments were performed,
detailed in table~\ref{runProperties_tab}. 
First, at
$t=1000$ in our standard 2D run (see section~\ref{2drun_sec}), we have
frozen the temperature and calculated the subsequent evolution of the vortex 
using a locally isothermal equation of state (model
$\textrm{STD2D}$). We found the vortex remained at its formation
location. Immediately, it is clear that the gas thermodynamics is crucial to the
migration process. 

To further test that idea, we performed a set
of simulations that reproduce the setup described by
\citet{2010ApJ...725..146P}. Within an isothermal and inviscid 2D disk, we
initialized a strong vortex by introducing a vortical velocity
perturbation. For three different exponents of the background density
radial profile (models $\textrm{PLP1}$, $\textrm{PLP2}$ and
$\textrm{PLP3}$), we measured three vortex migration velocities
that are in agreement with the results reported by \citet{2010ApJ...725..146P}. In
particular, the zero pressure-gradient case (model $\textrm{PLP3}$) is
almost neutral in the sense that the migration rate of the vortex is
vanishingly small.

 Next, in order to interrogate the role of the
thermodynamics, we relaxed the assumption of isothermality and
modified the  cooling function ${\cal L}$ in Eq.~(\ref{energy_eq}) so
that it takes the form
\begin{equation}
\cal L =
\begin{cases}
  (T-T_b/2)/\tau_v \, \, \, \,
  \textrm{if} \, \, \, \bb{\delta \omega}.\bb{e_z} < -1\\ 
  (T-T_b)/\tau_d \, \, \, \, \, \, \, \, \, \textrm{otherwise}.
\end{cases} 
\label{eq:PLPCOOL}
\end{equation}
$T_b$ is the gas background temperature, averaged
azimuthally\footnote{The second part of the cooling function in
Eq.~(\ref{eq:PLPCOOL}) might look unnecessary. Omitting that part, we
found that the disk was cooling down as a whole. This is because the
density waves launched by the vortex create large areas in the disk
where the vorticity is 
large. Such areas cool down as a result of the vorticity dependent
cooling function despite not being associated with the vortex itself,
an artefact that is taken care of by the second part of the cooling
function}. This is model $\textrm{PLPCOOL}$, and its piece-wise
cooling law, by depending on the strength of vorticity, forces the
vortex to possess a different temperature than its surrounding. 
If we take $\tau_v=10$ orbits and $\tau_d=1$ orbit, the vortex
is cooler than its surroundings which should relax to the initial
temperature 
profile (see top panel in
figure~\ref{bandeau_PLPCOOL}).
In this case, the vortex stays at its initial position, showing that a
change of the bulk temperature of the vortex is not enough by itself
to affect its migration. We then introduce an azimuthal asymmetry in
the vortex cooling by further modifying the cooling function: 
\begin{equation}
\cal L_{\pm}=
\begin{cases}
  (T-T_b(1 \pm \frac{\bb{v}\cdot\bb{e_R}}{2|\bb{v}\cdot\bb{e_R}|}))/\tau_v \, \, \, \,
  \textrm{if} \, \, \, \bb{\delta \omega}.\bb{e_z} < -1\\ 
  (T-T_b)/\tau_d \, \, \, \, \, \, \, \, \, \textrm{otherwise}.
\label{VortCoolAsym}
\end{cases} 
\end{equation}
The resulting relative temperature perturbations differ widely as
illustrated on the middle and bottom panels of
figure~\ref{bandeau_PLPCOOL}. With this new cooling 
prescription, we found significant inward or outward migration of the
vortex depending on the sign in Eq.(\ref{VortCoolAsym}), despite the
vanishing radial pressure gradient. If the vortex is hotter in the
region downstream of its core (in the sense of the background rotation)
than the surrounding gas while being cooler in the upstream region
(case $\textrm{PLPCOOL+}$), the vortex migrates inward. In the
opposite situation (case $\textrm{PLPCOOL-}$), the vortex migrates
outward with the same velocity. The migration speed is significant and
in fact comparable in magnitude (to less than a factor of two) to the
isothermal case when $p=-1.5$ (see table~\ref{runProperties_tab}). 

We conclude from this series of experiments that
diabaticity can play an important role in vortex migration. 
In particular, an azimuthal temperature gradient may help the
  vortex to overcome pressure gradients that would 
otherwise hold it in place.
The detailed and quantitative understanding of these effects is
largely beyond the scope of this paper and requires additional more specialised
simulations. However, it seems likely that a migration mechanism is at
work in our simulations that differs from that discussed in earlier
work. 
It is likely that the new effect is associated with
the baroclinic term in the vorticity equation, which vanishes
in the barotropic flow
of model PLP3 but is nonzero in model $\textrm{PLPCOOL}\pm$. This constitutes the
only difference between the two experiments.
We speculate that the
cooler temperatures in our vortex solutions
and the sharp temperature gradient at its surface would mean that the
vortex circulation is modified, with fluid parcels having to turn
abruptly. This could modify the position of the sonic lines and, along
with the modified vortex shape (see the black lines on
figure~\ref{bandeau_PLPCOOL}), maybe strengthen their asymmetry and 
trigger radial migration.

In addition, the
importance of thermodynamics effects is consistent with the
differences we find between the 2D and 3D simulations, particularly
when it comes to comparing the migration velocities of the
vortex. Obviously, the thermal history of the vortices in these
simulations differ on account of the two distinct heating processes
 (turbulent and ohmic dissipation in
the 3D case, viscous dissipation in the 2D simulation). The fact that
the different baroclinic terms might result in different vortensity
evolution is partly supported by the left panels of
figure~\ref{bandeau} that shows the different vorticity field
distribution within a single vortex in the two cases.

\subsection{The case of the static dead zone phase}

\begin{figure}
\begin{center}
\includegraphics[width=7cm]{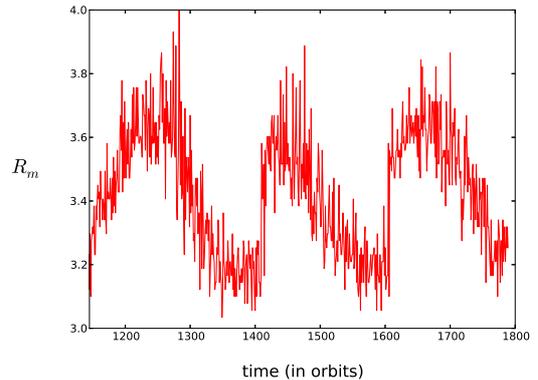}
\caption{Radial position of the density maximum in the 3D run
  during the static dead zone step.} 
\label{static_migration}
\end{center}
\end{figure}

We finally end the discussion by noting that our findings
naturally explain the difference between the static 
dead zone step and the self-consistent step. Indeed, a vortex is
observed to grow in the former. The question is: why does not this
vortex migrate into the dead zone? We have checked this issue and we
have found that, in fact, vortices do migrate inward in this case as
well. This is illustrated by figure~\ref{static_migration}. The
difference in this case is that the vortex migration range is much
smaller than in the 3D case: the vortex disappears upon reaching $R
\sim 3.1$ (as opposed to $\sim 2.7$ in the 3D case). This is because
the resistivity is not a function of temperature in this case, but a
function of position only: as the vortex penetrates into the active
zone, its resistivity drops to zero with renders the flow vulnerable
to MRI induced velocity fluctuations. As a result, vortices formed
during the static dead zone step have a smaller lifetime (of order
$100$ orbits, as opposed to the $300$ orbits measured during the
self-consistent step) when they enter the active zone.

\section{Conclusion}
\label{ccl}

In this paper, we have performed a non-ideal MHD simulation of the
region centred around the dead-zone inner edge of a protoplanetary
disk. In accordance with previously published work
\citep{lyra&maclow12}, we find that a vortex forms at the dead zone
inner edge. Our simulation reveals that the vortex is not affixed to
the pressure maximum as one would expect from the results of
\citet{2010ApJ...725..146P} but migrates inward. It penetrates into
the active zone where it is gradually destroyed by turbulent motions. 
A few orbits later a new vortex forms at the interface and follows the
same evolution, thereby creating what we have called a ``vortex
cycle'': formation, migration, and disruption follow each other and
comprise a quasi-periodic disk evolution. 

We find that the vortex cycle does not exist in isothermal
simulations of the dead zone inner edge. This is because the vortex
stays at its formation location without migrating inward.
In fact, we have
been unable to design any test in which a vortex moves in a
uniform pressure background when its temperature is equal to that of
its surroundings. Vortex migration seems to occur only if the
vortex temperature is different from (and maybe even free to evolve
independantly of) the temperature of its environment.
A systematic study and a detailed understanding of vortex
migration rates in such specific environments is needed.
 Indeed, we caution the reader that simulations
using a different cooling function may exhibit cycles with very
different properties. Moreover, we completely neglected heat
diffusion in the present paper, and particularly radiative
diffusion. Even if vortices forming in a optically thick region such
as the dead zone inner edge are likely to be cooler than the turbulent
region in which they will penetrate, radiative diffusion will act as a
heating source that will help decrease the difference between the
vortex temperature and the temperature of its surroundings.
If the heat diffusion timescale is shorter than the vortex cycle,
the ionization threshold will be crossed at some point. This will 
quickly activate the magneto-elliptic instability
\citep{2004ApJ...609..301L,2012JFM...698..358M}, the growth rate of 
which (about one local orbit) is much faster than the vortex
cycle timescale, and will disrupt the vortex.
An accurate measurement of the cycle period and the vortex migration rate
should be done using simulations including radiative transfer that
properly account for the vortex thermodynamics.

An additional limitation of our work comes from geometry itself. The
3D simulation uses the cylindrical approximation. Taking into account
the disk vertical structure \citep{2013A&A...559A..30R,
  2012A&A...542A...9M, 2009A&A...498....1L} will affect both the
vortex properties and the shape of the dead/active interface. Along
the same lines, the magnetic configuration is also known to influence
the development of vortices
\citep{2011A&A...527A.138L,2009ApJ...702...75Y,2013MNRAS.429.2748Y}. For
example, the presence of a net vertical flux in the inner parts might
change the results presented here in surprising ways.

Before closing this paper, we speculate on the possible consequences
of the vortex cycle for the dynamics of dust particles in the disk. As
we already discussed, vortices such as those we see in our simulations
are known to concentrate dust grains and help planet formation
processes. However, we have shown here that vortices do not stay at
their formation location but migrate inward, most likely carrying
the dust particles they captured at the density bump. If the
particles are still small once they are released by the disrupted vortex,
they might continue to migrate inward due to the gas friction. The
vortex cycle would then help the dust to pass across the pressure bump
barrier. From this discussion, a number of questions arise: 
\begin{enumerate}
\item What dust concentration can the vortex achieve?
\end{enumerate}
The 3D bi-fluid simulations of \citet{meheutetal12} have shown that
the  dust-to-gas ratio may change from $0.01$ to $1$ inside a vortex
in three local orbits. This is much shorter than the cycle period,
suggesting that vortices formed at the dead zone inner edge would trap
the entire content of dust initially present in the bump. 
\begin{enumerate}
\setcounter{enumi}{1}
\item Can dust grains embedded in the vortex become large enough
  as a result of collisions alone that
  friction becomes negligible at the time of vortex disruption? 
\end{enumerate}
According to 1+1D models of dust growth (e.g.\ Brauer et al.~2008a), 
the dust growth rate in laminar
regions is mainly due to differential settling. Such a growth
timescale can be taken as a proxy for the typical time required to
grow centimetre size  particles into metre size bodies within the
vortex. It amounts to about a thousand years which is much longer than
the cycle period of a few hundred years. This seems to suggest that
particles transported by a vortex across the bump barrier would not
grow significantly over one cycle. Particles would then
quickly drift toward the central star and take no part in the process of
planetesimal formation. A fast mechanism is needed to prevent a 
loss of the disk solid content. Given the large dust--to--gas ratio
that are likely to be found inside vortices, such fast processes
could be the streaming instability
\citep{2007ApJ...662..613Y,2007ApJ...662..627J}, gravitational
collapse \citep{2008A&A...491L..41L} or a combination of 
both phenomena \citep{2007Natur.448.1022J}. However, we caution the
reader that gas choatic motions in the vortex resulting from sound waves may
disrupt embryos formed by those mechanisms. A detailed study of the
outcome of these instabilities is needed.



For such a nonlinear and complex problem, it is difficult to go beyond
these simple qualitative statements. Clearly, self-consistent
simulations including the vortex cycle phenomenon along with dust
dynamics (including dust growth) are needed if we want to make any
quantitative statement regarding the fate of dust particles at the
dead-zone's inner edge. As we already argued, such simulations will
also have to properly include radiative effects since the vortex
migration is sensitive to the gas thermodynamics. Such multi--fluid
radiative MHD simulations are, of course, very resource demanding with
present day computing resources. 

\section*{ACKNOWLEDGMENTS}
The authors acknowledge a useful report from Dr Wladimir LYRA that
significantly strengthen the results presented in this paper. JF, SF
and HM acknowledge funding from the European Research Council 
under the European Union's Seventh Framework Programme (FP7/2007-2013)
/ ERC Grant agreement n° 258729. HL acknowledge support via STFC grant 
ST/G002584/1. The simulations presented in this paper were granted
access to the HPC resources of Cines under the allocation x2013042231
and x2014042231 made by GENCI (Grand Equipement National de Calcul
Intensif).

\bibliographystyle{aa}
\bibliography{main}

\end{document}